\documentclass[fleqn,usenatbib]{mnras}

\usepackage[T1]{fontenc}
\usepackage{ae,aecompl}

\usepackage{graphicx}	
\usepackage{amsmath}	
\usepackage{amssymb}	

\newcommand{\kms}{\,km\,s$^{-1}$}                          

\title[Conditions in the WR 140 WCR]
      {Conditions in the WR\,140 wind-collision region revealed by the 1.083-$\mu$m He\,{\sc i} line profile.}

\author[P. M. Williams et al.]{
Peredur M. Williams,$^{1}$\thanks{E-mail: pmw@roe.ac.uk}
Watson P. Varricatt,$^{2}$
Andr\'e-Nicolas Chen\'e,$^{3}$
\newauthor
Michael F. Corcoran,$^{4,5}$
Ted R. Gull,$^{6}$
Kenji Hamaguchi,$^{4,7}$
Anthony F. J. Moffat,$^{8}$
\newauthor Andrew M. T. Pollock,$^{9}$
Noel D. Richardson,$^{10}$
Christopher M. P. Russell,$^{11}$
\newauthor Andreas A. C. Sander,$^{12}$
Ian R. Stevens$^{13}$
Gerd Weigelt$^{14}$
\\
$^{1}$Institute for Astronomy, University of Edinburgh, Royal Observatory, Edinburgh EH9 3HJ, UK\\
$^{2}$Institute for Astronomy, UKIRT Observatory, 640 North A`oh\=ok\=u Place, Hilo, Hawaii 96720, USA\\
$^{3}$Gemini Observatory, Northern Operations Center, 670 North A`oh\=ok\=u Place, Hilo, HI 96720, USA\\
$^{4}$CRESST II and X-ray Astrophysics Laboratory NASA/GSFC, Greenbelt, MD 20771, USA\\
$^{5}$Department of Physics, Institute for Astrophysics and Computational Sciences, The Catholic University of America,\\
  Washington, DC 20064, USA \\
$^{6}$Astrophysics Science Division, NASA/GSFC, Greenbelt, MD 20771, USA \\
$^{7}$Department of Physics, University of Maryland, Baltimore County, 1000 Hilltop Circle, Baltimore, MD 21250, USA \\
$^{8}$D\'epartement de physique and Centre de Recherche en Astrophysique du Qu\'ebec (CRAQ), Universit\'e de Montr\'eal,\\
 C.P. 6128, Succ. Centre-Ville, Montr\'eal, Qu\'ebec, H3C 3J7, Canada \\
$^{9}$Department of Physics and Astronomy, University of Sheffield, Hounsfield Road, Sheffield S3 7RH, UK \\
$^{10}$Department of Physics, Embry-Riddle Aeronautical University, Prescott, AZ 86301, USA \\
$^{11}$Department of Physics and Astronomy, University of Delaware, Newark, DE 19716, USA \\
$^{12}$Armagh Observatory and Planetarium, College Hill, Armagh BT61 9DG, UK \\
$^{13}$School of Physics and Astronomy, University of Birmingham, Edgbaston, Birmingham, B15 2TT, UK \\
$^{14}$Max Planck Institute for Radio Astronomy, Auf dem H\"ugel 69, D-53121 Bonn, Germany \\
}
\date{Accepted 2021 February18.
      Received 2021 January 27;
      in original form 2020 November 30}

\pagerange{\pageref{firstpage}--\pageref{lastpage}}
\pubyear{2020}

\begin{document}

\maketitle

\label{firstpage}

\begin{abstract}

We present spectroscopy of the P~Cygni profile of the 1.083-$\mu$m \ion{He}{i} 
line in the WC7 + O5 colliding-wind binary (CWB) WR\,140 (HD 193793), 
observed in 2008, before its periastron passage in 2009, and in 2016--17, 
spanning the subsequent periastron passage. Both absorption and emission 
components showed strong variations. The variation of the absorption component 
as the O5 star was occulted by the wind-collision region (WCR) sets a tight 
constraint on its geometry. While the sightline to the O5 star traversed the 
WCR, the strength and breadth of the absorption component varied 
significantly on time-scales of days. 
An emission sub-peak was observed on all our profiles. The variation of its 
radial velocity with orbital phase was shown to be consistent with formation 
in the WCR as it swung round the stars in their orbit. 
Modelling the profile gave a measure of the extent of the sub-peak forming 
region. In the phase range 0.93--0.99, the flux in the sub-peak increased 
steadily,  approximately inversely proportionally to the stellar separation, 
indicating that the shocked gas in the WCR where the line was formed was 
adiabatic. After periastron, the sub-peak flux was anomalously strong and 
varied rapidly, suggesting formation in clumps down-stream in the WCR. 
For most of the time, its flux exceeded the 2--10-keV X-ray emission, 
showing it to be a significant coolant of the shocked wind. 

\end{abstract}


\section{Introduction}
\label{Sintro}

The collision of the hypersonic winds from the Wolf-Rayet (WR) and O stars in 
a massive binary system gives rise to a rich variety of phenomena observed 
from the radio to X-rays. 
Strong shocks are formed where the winds collide, leading to acceleration of 
particles, and heating and compression of the winds behind the shocks. 
The shock-compressed wind flows within the wind-collision region (WCR) and 
can sometimes be observed through the appearance of orbital phase-dependent 
`sub-peaks' on emission lines and, in some cases involving WC-type stars, 
the formation of carbon dust.

The subject of the present study is the WC7+O5\footnote{Different spectral 
subtypes have been assigned over the years. The most complete classification 
is WC7pd+O5.5fc by \citet{Remi}, who employed spectral disentangling to 
classify the O star, but we use WC7 and O5 for brevity here.} 
colliding-wind binary (CWB) system WR\,140 (HD 193793), which has become an 
archetypal system on account of the strong variations in its radio, X-ray 
and infrared emission -- the last caused by episodic dust formation --  all 
phase-locked to its highly elliptical orbit \citep[e.g.][]{W90}. 
Variation of the profile of its 5696-\AA\ \ion{C}{iii} emission line near 
the time of the 1993 periastron was first reported by \citet{Hervieux}. 
Around the following (2001) periastron, \citet{MM03} studied variations in 
the profiles of the C\,{\sc iii} and 5896-\AA\ He\,{\sc i} lines, while 
\citet*{VWA} studied those of the 1.083-$\mu$m He\,{\sc i} line. 
Previous spectroscopy of WR\,140 by \citet*{VreuxNIR}, \citet*{EWW} 
and \citet{W125} had shown the 1.083-$\mu$m emission-line profile to have 
a flat top, characteristic of formation in the asymptotic region of the 
WR wind, with no evidence of a sub-peak which could be formed in a WCR -- 
but these observations happened to have been taken at orbital phases 
(0.56, 0.41  and 0.82 respectively) far from periastron passage. 
Nearer periastron, between phases 0.96 and 0.02, \citet{VWA} observed 
conspicuous sub-peaks on the 1.083-$\mu$m line which, like those observed 
in the optical by \citet{MM03}, shifted during orbital motion consistently 
with the changing orientation of the WCR and the flow of the emitting 
material along it. They also showed that the maximum radiative flux in the 
1.083-$\mu$m He\,{\sc i} sub-peaks was greater than the 2--6 keV X-ray flux 
near the 1985 periastron or the 1--10 keV flux observed soon after the 1983 
periastron, and was therefore a significant source of cooling of the shocked 
WC7 wind. 
Of course, the O5 stellar wind is also shocked in the WCR and we have to 
consider whether there is a contribution to the 1.083-\micron\ sub-peak from 
its helium. The O+O CWBs have yet to be surveyed for 1.083-\micron\ sub-peak 
emission, but the spectroscopic survey of the 1-\micron\ region in OB stars 
by \citet{CH1083} finds no 1.083-\micron\ line emission in O5--O6 supergiants, 
including the CWB Cyg~OB2~9 \citep{CygOB9CWB} and the binary Cyg~OB2~11 
\citep{CygOB11SB}. Although we cannot rule out a contribution from the O5 
wind in WR\,140, these results suggest that the shocked O5 wind does not 
contribute to the sub-peak emission, or the undisturbed O5 wind to the 
underlying emission profile.

Unlike the 5696-\AA\ \ion{C}{iii} line, the profile of the 1.083-$\mu$m line 
in many WR stars also has a strong absorption component formed primarily in 
the asymptotic region of the stellar winds and valuable for measuring the 
terminal velocities, e.g. \citet{EWRV}. 
On the other hand, the O5 component is not expected to provide absorption in 
the 1.083-$\mu$m line, judging from the spectra of the O5f and O6f stars 
observed by \citet{CH1083}, or the O4V((f)) star 9~Sgr observed by \citet{VWA}, 
which do not show P~Cygni profiles in their 1.083-$\mu$m lines. The contrast 
of the He\,{\sc i} absorptions through the WC and O5 stellar winds can 
therefore provide a valuable tool for mapping the winds and WCR. The stars 
are too close together \citep[$a$ = 9~mas,][]{Monnier140} for us to resolve 
them with the spectroscopic instrumentation, so the observed profile is the 
superposition of the profiles formed in the separate, parallel sightlines 
to the O5 and WC7 stars. 
The 1.083-$\mu$m spectra of \citet{VWA} showed the absorption component to 
increase significantly between observations made before and after 
periastron passage as our lines of sight to the stars passed mostly through 
the O5 star wind in the first spectra and subsequently through the He-rich wind 
of the WC7 star in the later data. 
This allowed them to set constraints on the opening angle of the cone used to 
approximate the WCR, depending on the (then unknown) orbital inclination.

As part of the multi-wavelength campaign to observe WR\,140 around the time 
of the 2009 periastron, further observations of the 1.083-$\mu$m line profiles 
were  made during 2008 from phase 0.93 to just before periastron. 
A strong sub-peak was present on all spectra, while a sudden increase of the 
absorption component near phase 0.99 was used to estimate the opening angle 
of the WCR cone. 
A preliminary account of that work was given by \citet*{WVA} but we undertook 
two further observing campaigns in 2016--17 to cover the following periastron 
passage in order to form a more complete picture of the evolution of the WCR 
at the most critical phases. 

The principal scientific goals are to use the variation in 1.083-$\mu$m profile 
as the orbit progressed to map the WCR and to compare the flux emitted in the 
emission sub-peaks on the 1.083-$\mu$m line, which occur over a wider range in 
phase than those on the optical lines, with the X-ray fluxes to study the 
cooling of the shocks. 
We had intended to compare the profiles of the 1.083-\micron\ and 5696-\AA\ 
sub-peaks observed contemporaneously to see if they formed in the same or 
different regions of the WCR, but this was thwarted by poor observing 
weather at the critical phases. 
Although our observations cover less then one-seventh of the orbital period, 
they cover over 80 per cent of the orbit in terms of angular motion, so great 
is the eccentricity. 

In this paper, Section~\ref{SObs} reports the collection of the data and 
Section~\ref{SResults} presents the results: beginning with an overview 
of the variation of the line profiles, followed by discussion of the 
absorption and sub-peak emission components. We discuss the results and 
relate them to studies at other wavelengths in Section~\ref{SComp} and 
summarise conclusions in Section~\ref{SConclude}.

\section{Observations}
\label{SObs}%

\begin{table*}
\centering
\caption{Log of observations with UIST on UKIRT ordered by phase: those made in 
2008 were taken using the 2-pixel slit and those in 2016 with the 4-pixel slit. 
Dates are UT, quoted to 0.1~d.
Also given are the EWs of the absorption component and the fluxes in the sub-peaks, 
followed by their flux-weighted central velocities.}
\label{TUIST}
\begin{tabular}{lcccc}
\hline
Date                 & phase  & absorption & sub-peak flux &   RVc          \\
                     &        & EW (\AA)   & ($10^{-14}$W m$^{-2}$) & (\kms) \\
\hline
2008 June 27.6     & 0.9304 & 3.1$\pm$0.2 &  2.5$\pm$0.2 & -1391$\pm$102  \\  
2008 July 30.5     & 0.9417 & 2.4$\pm$0.1 &  2.4$\pm$0.1 & -1431$\pm$101  \\
2008 August 5.4    & 0.9438 & 2.3$\pm$0.1 &  2.8$\pm$0.2 & -1407$\pm$102  \\
2008 August 5.5    & 0.9438 & 2.2$\pm$0.1 &  2.6$\pm$0.2 & -1413$\pm$102  \\
2008 August 22.4   & 0.9496 & 2.3$\pm$0.1 &  2.7$\pm$0.2 & -1454$\pm$104  \\
2016 August 10.4   & 0.9555 & 2.3$\pm$0.1 &  3.3$\pm$0.3 & -1462$\pm$56   \\ 
2016 August 25.3   & 0.9600 & 2.4$\pm$0.1 &  4.0$\pm$0.2 & -1476$\pm$55   \\ 
2016 September 4.4 & 0.9634 & 2.3$\pm$0.1 &  4.3$\pm$0.2 & -1455$\pm$56   \\
2016 September 19.4& 0.9686 & 2.2$\pm$0.1 &  4.6$\pm$0.3 & -1454$\pm$58   \\
2008 December 8.2  & 0.9869 & 2.9$\pm$0.1 &  7.5$\pm$0.2 & -1080$\pm$102  \\
2008 December 19.2 & 0.9907 & 8.5$\pm$0.1 &  8.5$\pm$0.3 &  -697$\pm$115  \\  
2008 December 20.2 & 0.9910 & 7.8$\pm$0.1 &  8.9$\pm$0.2 &  -712$\pm$102  \\ 
2008 December 21.2 & 0.9914 & 8.4$\pm$0.1 &  8.7$\pm$0.2 &  -645$\pm$102  \\ 
2008 December 22.2 & 0.9917 & 9.3$\pm$0.1 &  8.2$\pm$0.4 &  -639$\pm$104  \\ 
2008 December 23.2 & 0.9920 & 9.3$\pm$0.1 &  8.6$\pm$0.4 &  -613$\pm$104  \\ 
2008 December 24.2 & 0.9924 & 8.3$\pm$0.1 &  9.0$\pm$0.4 &  -563$\pm$103  \\
\hline
\end{tabular}
\end{table*}

\begin{table*}
\centering
\caption{Log of observations with GNIRS, giving EWs of the absorption component, 
fluxes and flux-weighted central velocities. The absorption EWs and sub-peak 
fluxes observed after periastron take account of the contribution of dust 
emission to the continuum as described in the text.}
\label{TGNIRS}
\begin{tabular}{lrccc}  
\hline
Date               &  phase & absorption & sub-peak flux     &     RVc     \\
                   &        & EW (\AA) & ($10^{-14}$W m$^{-2}$) & (\kms)   \\
\hline
2016 December 15.2 & 0.9986 &  9.3$\pm$0.3 &    22.3$\pm$0.3 & -148$\pm$40 \\  
2016 December 22.2 & 0.0010 & 10.0$\pm$0.2 &    22.6$\pm$0.9 &  482$\pm$84 \\  
2016 December 23.1 & 0.0013 & 10.4$\pm$0.3 &    19.1$\pm$0.4 &  710$\pm$99 \\  
2016 December 24.0 & 0.0016 & 10.5$\pm$0.2 &    18.6$\pm$0.5 &  843$\pm$30 \\  
2016 December 27.2 & 0.0028 & 10.6$\pm$0.2 &    17.7$\pm$0.5 &  950$\pm$55 \\  
2017 March 23.6    & 0.0326 &  4.6$\pm$0.2 &    18.5$\pm$0.5 &  440$\pm$30 \\  
2017 March 26.7    & 0.0337 &  7.2$\pm$0.1 &    14.5$\pm$0.4 &  416$\pm$30 \\  
2017 March 27.7    & 0.0340 &  6.5$\pm$0.1 &     9.0$\pm$2.0 &  512$\pm$30 \\  
2017 March 28.6    & 0.0343 &  7.2$\pm$0.2 &    13.2$\pm$0.6 &  389$\pm$30 \\
2017 March 30.6    & 0.0350 &  6.9$\pm$0.2 &    15.2$\pm$0.5 &  418$\pm$30 \\
2017 April 24.6    & 0.0437 &  6.4$\pm$0.2 &    16.2$\pm$0.6 &  333$\pm$30 \\  
2017 June 17.6     & 0.0623 &  6.5$\pm$0.1 &    10.3$\pm$0.4 &  495$\pm$30 \\  
2017 June 18.6     & 0.0626 &  6.7$\pm$0.1 &    10.5$\pm$0.5 &  308$\pm$40 \\  
2017 June 19.6     & 0.0630 &  6.3$\pm$0.1 &     6.4$\pm$0.7 &  288$\pm$60 \\  
\hline
\end{tabular}
\end{table*}

The 2008 observations were made with the United Kingdom Infrared Telescope 
(UKIRT) on Mauna Kea, Hawaii, using the 1--5 micron UKIRT Imager Spectrometer 
(UIST) \citep{UIST} in programme U/08B/17. The short-$J$ grism and 4-pixel 
slit gave a resolution of 200~\kms. 
Observations generally comprised 12 integrations of 30~s, and spectra of the 
F5V star BS 7756 were observed at comparable airmass to WR\,140 to correct for 
telluric absorption features, which are signifcant in this wavelength region. 
Wavelength calibration was performed  an argon lamp.

The first of our observations in 2016 were taken at phases close to conjunction 
(O5 star in front), also using UIST on UKIRT (programme U/16B/UA10). 
For these, the slit width set to 2 pixels, giving a higher resolution of 
100~\kms, and the A2V star BS 7769 was used as an additional telluric standard. 
 
The spectra were not flux-calibrated at the time of observation but flux 
calibration is provided via the continuum determined from the $r$ and $J$ 
photometry of the stellar wind. The level is taken to be constant during these 
observations because IR photometry, including observations in 2008, show that 
that the dust emission in the $J$ band from the previous dust formation 
episode was no longer observable after phase 0.25 \citep{Dust140,Taranova140}. 
The stellar-wind continuum flux density at 1.08~$\mu$m, corrected for 
interstellar reddening, was 
found to be $F_{\lambda}$ = 4.33 $\times$ 10$^{-11}$~Wm$^{-2}\mu$m$^{-1}$. 
A log of the UIST observations is given in Table \ref{TUIST}, together with 
equivalent widths of the absorption component and parameters of the emission 
subpeak to be discussed below. 
The phases were calculated using the ephemeris \citep{Thomas21} for periastron:

\begin{equation}
T_0 (MJD) = 60636.23 + 2895.00E.                                  
\end{equation}

Further observations in 2016--17 close to, and shortly after, periastron were 
made with Gemini North, also on Mauna Kea, using the Gemini Near-InfraRed 
Spectrograph \citep[GNIRS,][]{GNIRS} in programmes 2016B-Q-49 and 2017A-Q-13. 
GNIRS was used in long-slit mode with the 110.5~line~mm$^{-1}$ grating in the 
6th order (X band), short blue camera and 2-pixel slit, giving a resolution 
of about 49~\kms, higher than those of the UIST spectra. 
Wavelength calibration was from an argon lamp. Each observation comprised 
eight integrations of 5~s, sometimes split over 2--3 co-adds. To correct for 
telluric aborption lines, spectra of the A1V stars HIP 99893 or HIP 103108 
were observed at comparable airmass. Besides the strong Paschen $\gamma$ 
line at 1.0941 $\micron$, the A1V stars have significant C\,{\sc i} lines 
at 1.0687, 1.0694, 1.0710 and 1.0732 $\mu$m which were corrected for. 
Unfortunately, the HIP~103108 calibration spectrum on December 15 was observed 
in the wrong wavelength region and it was necesary to use a spectrum from 
another night and shift its wavelength scale to cancel, as far as possible, 
the telluric lines in the WR\,140 spectrum. 
The GNIRS spectra were not flux-calibrated at the time of observation and 
again our photometrically derived 1.08-\micron\ continuum was used. 
For most of these observations it is necessary to allow for the additional 
contribution of dust emission to the continuum. This had the effect of 
diluting the line emission and absorption in our spectra. The additional flux 
was determined from the $J$ band photometry and assumed to be dust emssion 
described by that of amorphous carbon grains at a temperature of 1100~K 
\citep{Dust140}, allowing calculation of the 1.08-\micron\ flux. 
From zero in the December 15 spectrum, the dust contribution at 1.08~$\micron$ 
grows to add 3.7~per cent of the stellar wind continuum by the December 27 
observation, 7 per cent during the March-April series and 6 per cent during 
the June observations.
A log of the GNIRS observations is given in Table~\ref{TGNIRS}, where the 
equivalent widths of the absorption components and the fluxes in the sub-peaks 
have been adjusted to correct for the contribution by dust emission to the 
continuum.
As an overview of the movement of the sub-peaks, their flux-weighted mean 
radial velocities (RVs) (as used by \cite{Remi} for the 5696-\AA\ sub-peaks) 
were measured and are also given in Tables~\ref{TUIST} and \ref{TGNIRS}. 

\section{Results}
\label{SResults}

\begin{figure}               
\centering
\includegraphics[width=8.5cm]{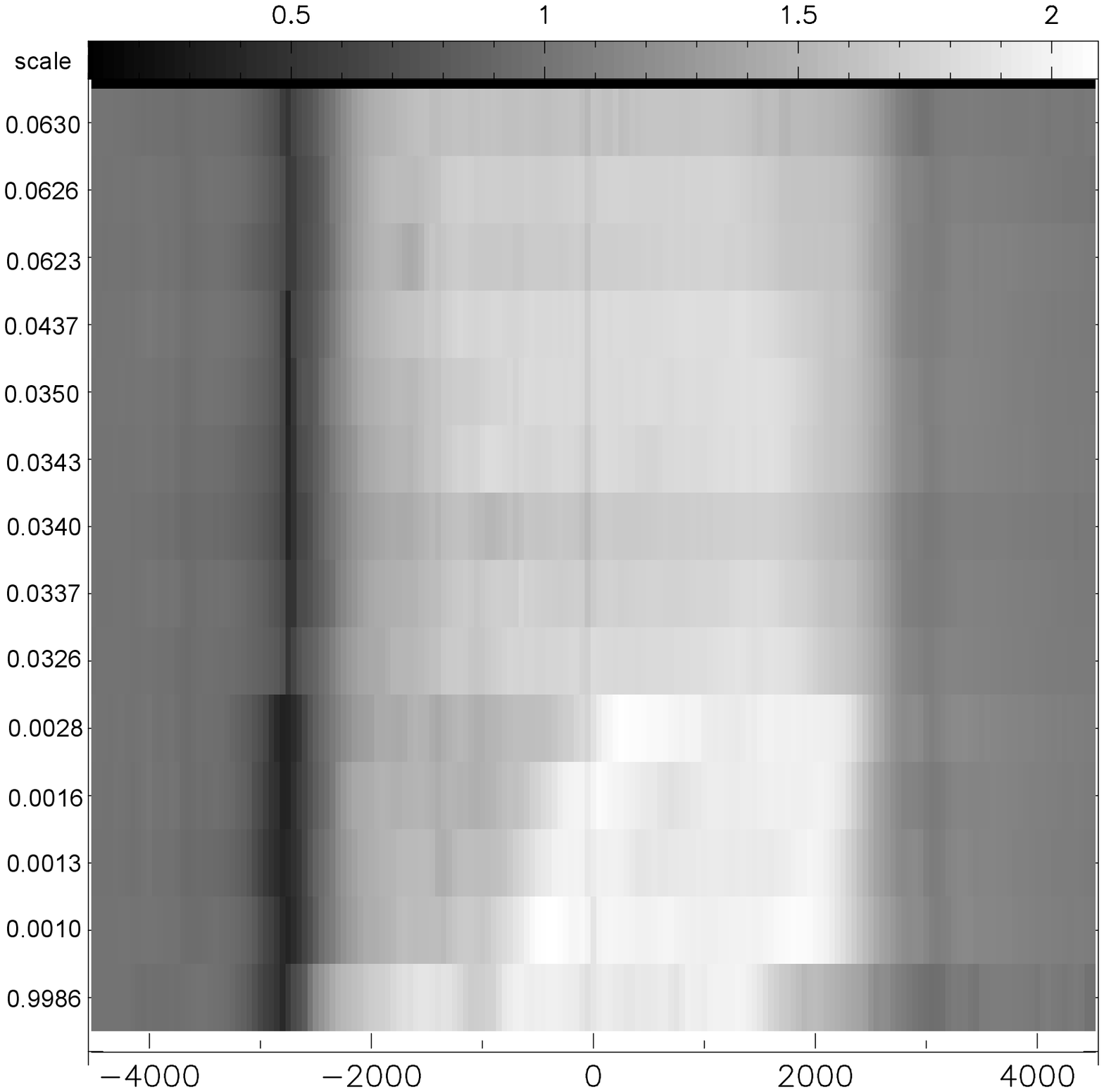}                       
\includegraphics[width=8.5cm]{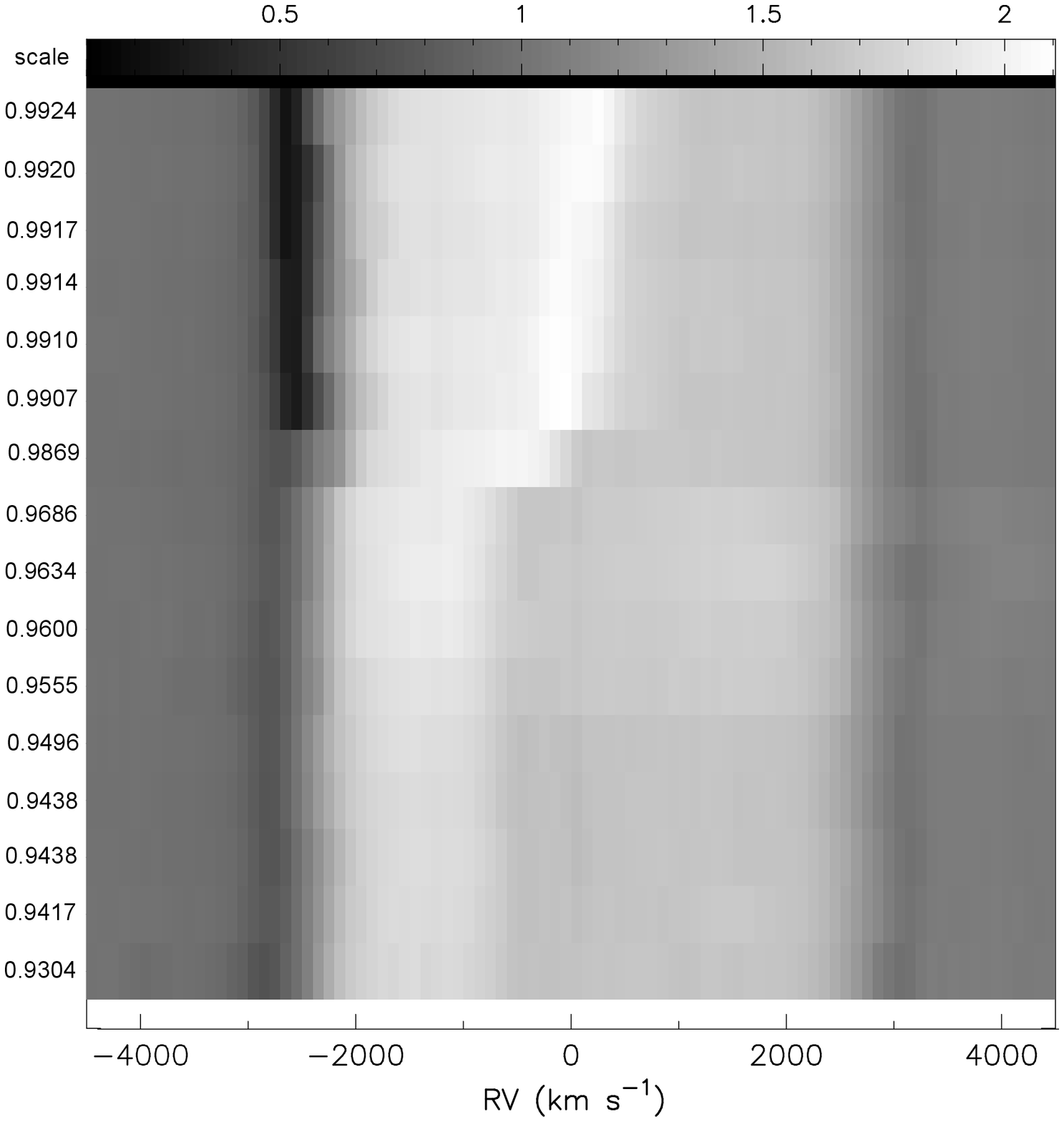}          
\caption{Dynamic spectra of the 1.083-$\mu$m profile before and after 
periastron, sequenced by orbital phase starting from $\phi = 0.93$
reading upwards, top panel from Gemini/GNIRS and lower panel from UKIRT/UIST. 
Flux scales are given on top of each panel. The data are not evenly spaced in 
phase and some abrupt changes reflect gaps in the coverage as can be 
seen from the labels on the y-axis. As described in the text, the 
spectra were observed at different resolutions.}
\label{Fdynamic}
\end{figure}                 

\subsection{Evolution of the line profile}
\label{SEvolve}

\begin{table}                             
\centering
\caption{{\bf Benchmark phases in the orbit and that of the beginning of the  
sharp rise in absorption discussed below.}}
\label{Tphases}
\begin{tabular}{ll}
\hline
Phase  & phenomenon \\  
\hline
0.0030 & Conjunction, O5 star behind \\
0.0360 & First quadrature \\
0.9554 & Conjunction, O5 star in front \\
0.9966 & Second quadrature \\
0.986  & beginning of sharp rise in absorption. \\ 
\hline
\end{tabular}
\end{table}

A synoptic view of the evolution of the line profile is provided by the dynamic 
spectra presented in Fig.\,\ref{Fdynamic}. The UKIRT/UIST and Gemini/GNIRS 
spectra conveniently fall into two sets separated by orbital phase and 
spectral resolution. The UKIRT/UIST (lower panel) spectra start at 
phase 0.9305, shortly before conjunction (O5 star in front, $\phi$ = 0.9554), 
and continue to $\phi$ = 0.9925, shortly before periastron. 
Given in Table~\ref{Tphases} for reference are the phases of the 
conjunctions and quadratures calculated using the the values of orbital 
eccentricity, argument of periastron (O5 star) and inclination determined 
by \citet{Thomas21}: $e = 0.8993$, $\omega = 47\fdg44$, $i = 119\fdg07$.
The profiles all show a sub-peak on the broad emission component, initially 
apparently stationary at the `blue' end (RV $\simeq$ --1420 km~s$^{-1}$) 
of the profile, and subsequently broadening and moving to longer 
wavelengths, it strengthened significantly between phases 0.9687 and 0.9867. 
We do not know the phase at which the subpeak first appeared; as noted above, 
it was not present in the profile \citet{W125} observed at phase 0.83. 
The absorption component in our early spectra appears roughly constant before 
suddenly increasing sharply in the 11 days between phases 0.9869 and 0.9907.
Thereafter, it remains strong and conspicuously variable in the approach to 
periastron (not reached in this sequence of observations as the source was setting), 
while the emission moves steadily to higher velocity.  

The Gemini/GNIRS spectra (upper panel of Fig.\,\ref{Fdynamic}) fall into three 
fairly concentrated sequences, in 2016 December, 2017 March--April and 
2017 June. The peak in the broad excess emission moved redward, especially 
between the first two spectra observed a week apart as the system went 
through periastron, and then continued moving redward to phase 0.0026 
(December 27). Owing to their higher resolution (49 \kms), the GNIRS spectra 
resolve the absorption component better than the UIST (100--200 \kms) 
spectra and, after phase 0.0324, show it to have a narrow core 
together with a variable, broad component. 
There is also a broad transient absorption feature near --1650~\kms\ in the 
2017 June 17 spectrum ($\phi$ = 0.0621) which had vanished by the following night. 
The broad emission sub-peak continued its movement to higher velocities 
in the first five spectra, until 2016 December 27; thereafter, it was broader 
and weaker in the subsequent observations from 2017 March 23.
The profiles are discussed below.

\subsection{The absorption component of the 1.083-$\mu$m He\,{\sc i} line profile}
\label{SAbs}

The equivalent widths (EW) of the absorption components measured from our UIST 
and GNIRS spectra by direct integration are given in Tables \ref{TUIST} and 
\ref{TGNIRS}. To ensure 
comparability of the results from the different instruments, the continuum 
for each measurement was determined by fits to the same wavelength regions, 
1.069--1.071 and 1.104--1.108~$\mu$m, while errors were estimated from 
repeated measurements varying the choice of the `blue' edge of the profile. 
The red edge of the profile is determined by the rising edge of the emission 
profile. A possible concern is that some emission from the sub-peak when it 
is at its shortest wavelength could overlap with and fill in part of the 
absorption component, thereby weakening both features. 
We cannot rule out this possibility given the absence of undisturbed wind 
emission shortward of the sub-peak. We note, however, that the absorption 
does not increase between the first and tenth spectra while the sub-peak was 
moving to longer wavelengths, giving an increase of over 300~\kms\ in RVc 
(Table~\ref{TUIST}; see also the profiles in Fig\,\ref{FUmodels7} below) 
when it would be less likely to fill in the absorption, suggests that this 
is not a significant effect.

In addition, the absorption components and sub-peak fluxes in the earlier UKIRT 
spectra observed by \citet{VWA}, \citet{EWW} and \citet{W125} were re-measured 
using, as far as possible, the same methodology as for the new data, including 
correction for the contribution by heated dust to the 1.08-$\mu$m continuum at 
the times of the 2001 March observations. 
The EWs and sub-peak fluxes from these observations are given in Table~\ref{TCGS4}. 

\begin{table}                             
\centering
\caption{Equivalent widths of the absorption component and sub-peak fluxes in 
2000--01 re-measured from the higher resolution (R = 4700) CGS4 spectra observed 
by \citet{VWA}, together with those from their UIST spectrum of 2003 and from 
the earlier CGS2 and CGS4 spectra observed by \citet{EWRV} and \citet{W125}. 
The EWs observed in 2001 also take account the contribution of dust emission to 
the continuum.}
\label{TCGS4}
\begin{tabular}{llrrl}
\hline
Date          & phase  & absorption    & sub-peak flux          & Inst. \\  
              &        & EW (\AA)   & ($10^{-14}$W m$^{-2}$) &      \\
\hline
2000 Oct 13   & 0.9583 & 3.0$\pm$0.1 &  3.3$\pm$0.3   &  CGS4 \\  
2000 Dec 25   & 0.9835 & 2.6$\pm$0.2 &  8.4$\pm$0.3   &  CGS4 \\  
2000 Dec 26   & 0.9838 & 3.2$\pm$0.2 &  8.4$\pm$0.3   &  CGS4 \\  
2001 Mar 18   & 0.0123 & 6.8$\pm$0.3 & 13.1$\pm$0.5   &  CGS4 \\  
2001 Mar 31   & 0.0168 & 6.5$\pm$0.2 & 13.2$\pm$0.5   &  CGS4 \\  
2003 May 24   & 0.2876 & 5.2$\pm$0.1 & 0.0            &  UIST \\  
1988 Jun 28   & 0.408 & 4.5$\pm$0.2 & 0.0            &  CGS2 \\  
1991 Oct 20   & 0.826 & 3.0$\pm$0.3 & 0.0            &  CGS4 \\  
\hline
\end{tabular}
\end{table}

\begin{figure}                               
\centering
\includegraphics[angle=270,width=9.0cm]{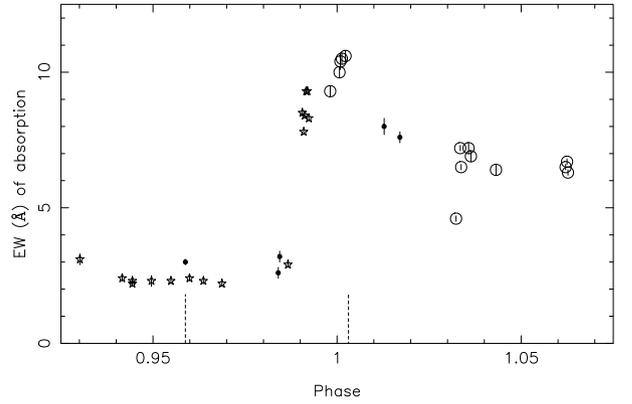}                        
\caption{Equivalent width (EW) of the absorption component from the UIST ($\star$), 
GNIRS ($\circ$) and \citet{VWA} ($\bullet$) plotted against phase. 
The error bars are $\pm1\sigma$.
Vertical broken lines mark conjunctions, the O5 star in front near phase 0.96 and 
the WC7 star in front just after periastron. The phases of the conjunctions 
are given in Table~\ref{Tphases}.} 
\label{FEWphase}
\end{figure}                                   

\begin{figure}                                
\centering
\includegraphics[width=8cm]{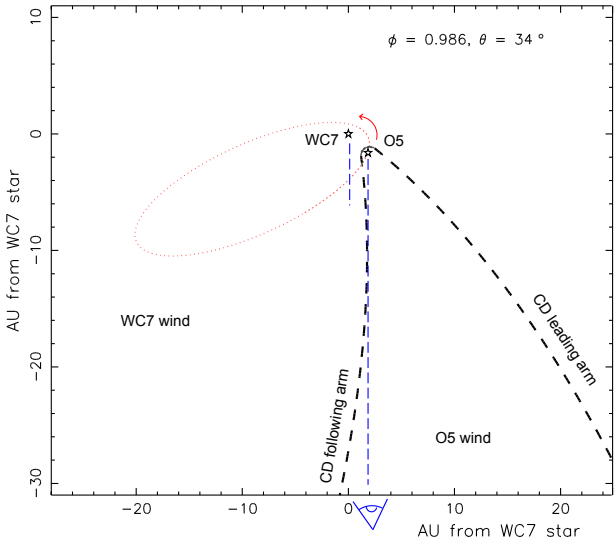}                        
\caption{Sketch of the WC7 and O5 stars and the WCR configuration projected on to 
the plane of the observer at phase 0.986 showing the beginning of the occultation
of the line of sight to the O5 star as it was intersected by the following arm  
contact discontuity (CD, heavy dashed line) moving counter-clockwise with the O5 
star in its orbit (red) in this representation. 
The curvature of the CD and WCR are a consequence of the orbital motion and 
the radial flow of the winds, and varies round the orbit as the transverse 
velocity varies. (The sightline to the WC7 star was crossed earlier but this 
was not expected to affect the extinction, most of which occurs closer to the 
WC7 star where its wind is densest.)}
\label{Foccult}
\end{figure}                                   

The EWs within $0.075P_{\textrm{orb}}$ of periastron are plotted against 
phase in Fig.\,\ref{FEWphase}. 
The highest EW is seen near conjunction, when the O5 star is furthest from us, 
but the most striking feature is the very sharp increase in the 11 days, 
2008 December 8--19, prior to this maximum caused by the passage of the 
following arm of the WCR across our sightline to the O5 star, illustrated in 
Fig.\,\ref{Foccult}. This increase must have begun at $\phi \simeq 0.986$, 
just before our December 8 observation,  (cf. Table~\ref{TUIST}) and is 
modelled in Section~\ref{Soccult} below.

As noted above, the observed absorption profile is the superposition of the 
profiles formed in the separate sightlines to the O5 and WC7 stars. That to 
the WC7 star always passes through at least part of the WC7 wind, that closest 
to that star, where the density is highest and most of the absorption occurs. 
Therefore, that component of the absorption is not expected to vary 
significantly round the orbit and we assume it to be constant. 
On the other hand, the extinction along the sightline to the O5 star through 
the WC7 wind varies systematically as the orientation and separation of the 
stars change with orbital motion and can be calculated given the orbit.
If the cavity in the WC7 wind blown by the O5 wind and WCR is large enough,  
and the orbital orientation favourable, the line of sight to the O5 star 
misses the WC7 wind for part of the orbit\footnote{Because of the orbital   
motion, the WCR wraps around the stars in a spiral and the sightlines may pass 
through wind of the WC7 star several times; but as the wind density in the 
outer turns is much lower, the extinction in only the first turn of the spiral 
will be considered here.}. 

This occurs around the time of conjunction when the O5 star 
and WCR are in front of the WC7 star, resulting in an interval 
(0.94 $<\phi<$ 0.97) of low absorption-component EW. 
At these phases, the cavity blown in the WC7 stellar wind by the WCR is 
oriented towards us, so that the O5 star is observed through its own wind only 
which, as suggested by the spectra of luminous O4--6 stars referred to above, 
is not expected to provide any extinction in the He\,{\sc i} line. 
The extinction profile observed at these phases is then just that formed in 
the sightline to the WC7 star through its own wind, diluted by the 
unextinguished continuum of the O5 star. 
We can estimate the dilution by noting that \citet{Monnier140} measured the 
WC7 star to be $1.37\times$ brighter than the O5 star in the $H$ band and 
comparing the continuum SEDs of WC7 and O star models in the 1.06--1.65-$\mu$m 
range. This yields a flux ratio near 1.02 (WC7/O5)\footnote{Extending this model 
to the visible, the flux ratio falls to 1.0, which compares with the range 
(1.37--0.5) found by \citet{Remi} from dilution of the spectral lines in one 
or other of the components compared with single stars of similar types.} in 
the region of the He\,{\sc i} line. 
Accordingly, we expect the EW of the absorption in the WC7-only profile to 
be around twice the EW observed around conjunction, i.e. 4.6\AA. This 
absorption will be present at all phases, together with that to the O5 star.

The variation of EW with phase shows significant scatter, greater than the 
observational uncertainties, on top of the expected smooth, orbitally dependent 
variation as the O5 star moves behind the WCR and into the denser WC7 wind. 
First we consider the orbitally dependent variation, then the origin 
of the scatter.

\subsection{Modelling the occultation of the O5 star}
\label{Soccult}  

The rise to maximum absorption (Fig.\,\ref{FEWphase}) appears to occur in two 
stages, first the sharp rise at $\phi \simeq 0.986$ and secondly the rise 
through periastron. Unfortunately, this result comes from two different cycles 
(2008 and 2016) and there is a gap in phase coverage, but we note a similar 
effect in the hardness ratio of the {\em RXTE} PCA data (Pollock et al., 
in preparation, Fig. 7), 
which also shows a pause in its increase near $\phi \simeq 0.995$, close 
to second quadrature on the way to maximum just after conjunction. 
The similarity of the X-ray and 1.083-\micron\ absorption variations, despite 
the fact that the X-ray source is not coincident with the O5 star but lies in 
the WCR presumably close to the shock apex, suggests that the second stage of the 
rise to maximum is caused by the movement of the O5 star and the X-ray source 
further into and behind the WC7 stellar wind. 

The 119\degr-orbital inclination prevents the O5 star from being eclipsed by 
the WC7 star, but there is a significant decrease in impact parameter, $p$, 
between our line of sight to the O5 star and the WC7 star. It is given by:

\begin{equation}
p = D \sin \psi 
\end{equation}

\noindent where $D$ is the separation of the stars and $\psi$ is the angle 
between our line of sight and the line of centres between the WC7 and O5 stars. 
This angle is found from

\begin{equation}
\cos(\psi) = -\sin(i)  \sin(f+\omega),
\label{psidef}
\end{equation}

\noindent where $f$ is the phase-dependent true anomaly and $i$ and $\omega$ are 
the orbital inclination and argument of periastron from \citet{Thomas21} quoted 
above. The impact parameter falls from $p = 0.118a$ at second quadrature to 
$p = 0.059a$ at conjunction, where $a$ is the 
length of the semi-major axis. Consequently, the sightline experiences 
significantly greater WC7 wind density and absorption between these phases. 
The increase in the extinction in the WC7 wind can be calculated using the 
relation for the X-ray extinction through a stellar wind as a function of 
orbit by \citet[Appendix]{W90}:

\begin{equation}
\tau  \propto \frac{\sec i}{r \cos(f+\omega) \surd \Delta} \left[ \arctan \left( \frac{-\surd \Delta}{\tan(f+\omega)\tan i}\right) \right]
\label{extn}
\end{equation}

\noindent where
\begin{equation}
\Delta = 1 + \tan^2(f+\omega) + \tan^2i, 
\end{equation}

\noindent $r$ is the distance from the intersection to the WC7 star and $f$, 
$\omega$ and $i$ are as above. The extinction increases by a factor of 3.3 
between these phases, but the influence on the observable EW of the \ion{He}{i} 
line is much smaller because of the presence of the extinction 
towards the WC7 star itself.

The first stage of increase in extinction, that near $\phi = 0.986$ is then 
interpreted as the passage of the edge of the WCR and WC7 wind across the 
sightline to the O5 star (Fig.\,\ref{Foccult}), which we now model. 
The WCR straddles the surface where the stellar wind momenta balance, the 
`contact discontinuity' (CD). Sufficiently far from the stars, and in the 
absence of orbital motion, the CD can be approximated by a cone 
(e.g. \citet{GirardWilson,EichlerUsov}), having a half angle, $\theta$, which 
is determined by the properties of the colliding stellar winds, particularly 
the ratio of their momenta:

\begin{equation}
\eta = \frac{(\dot{M}v_{\infty})_{\textrm{O5}}}{(\dot{M}v_{\infty})_{\textrm{WC7}}}.
\end{equation}

The relation between $\theta$ and $\eta$ has been studied for different 
conditions in the shocked material, including purely radiative and 
adiabatic shocks by, e.g., \citet{GayleyAngle}. In the present study, we 
will use the observations of the occultation near phase 0.99 to measure  
the angle $\theta$ directly and then consider $\eta$. 

To derive $\theta$ from the observations, we have to take account of two 
further effects, the inclination of the orbit and the twisting of the cone 
by the orbital motion of the stars (Fig.\,\ref{Foccult}). 
Seen from a non-zero inclination, the apparent opening angle of the cone 
will vary round the orbit, being equal to $\theta$ at quadratures only and 
smaller for most of the time, being reduced to zero if $\theta$ and the 
inclination are small enough. 
Writing $\theta^{\prime}$ for the half opening angle projected on to the 
observer's plane through the apex of the cone, it is related to $\theta$ by:

\begin{equation}
\cos \theta^{\prime} = \frac{\cos \theta}{\sin\,(\arccos\,(\cos i \sin (f+\omega)))}
\end{equation}

\noindent where $i$, $f$ and $\omega$ are as above. 
To model the twisting of the WCR by orbital motion, we require the recent 
history of the transverse velocity, $v_{\textrm{t}}$, of the O5 star in its 
orbit in the WC7 wind calculated from the orbital elements, and the expansion 
velocity, taken to be the terminal velocity \citep[2860 \kms,][]{WE2058} of 
the WC7 wind, which dominates the structure on account of its greater momentum 
\citep[cf. the consideration of the WR\,104 pinwheel by][]{Tuthill104ApJ}. 
An alternative position, that the expansion velocity is that of the slower 
wind \citep{ParkinPittard}, also points to the WC7 star because its terminal 
velocity is lower than that \citep[3100\kms,][]{Diah140} of the O5 star. 
The shapes of the leading and following arms of the CD in the observer's 
plane were calculated assuming the material to move ballistically at angles 
$\theta^{\prime}$ and $-\theta^{\prime}$ from the projected WC7--O5 axis, 
starting from the leading and following edges of the `rim' dividing the 
`shock cap' \citep{ParkinPittard}, the curved region of the CD between the 
stars, from the cone beyond the O5 star. The rim is perpendicular to the 
WC7--O5 axis and its radius was taken\footnote{The thin-shell models of 
\citet*{Canto} give values in the range 2.0--2.2 for this factor whereas 
\citet{EichlerUsov} give $\pi/2$. The exact choice within this range was 
found to make no difference to our modelling.} to be $2r_{\textrm{O5}}$, 
where $r_{\textrm{O5}}$ is the distance from the O5 star to the stagnation 
point of the WCR and is related to the separation of the stars, $D$, by

\begin{equation}
r_{\textrm{O5}} = \frac{\surd\eta}{1+\surd\eta}D.
\label{ErO5} 
\end{equation}
 
For each of a range of values of $\theta$, the system configuration and CD
were mapped for a sequence of orbital phases covering the observations.  
Because $v_{\textrm{t}}$ varies significantly around the orbit, so does the 
curvature of the CD, which depends on the recent history of $v_{\textrm{t}}$.

At each phase, the intersection of the line of sight to the O5 star with 
the boundaries, the `leading' and `following' arms (see Fig.\,\ref{Foccult}), 
of the CD were located and the distances to the WC7 star and absorption 
through its wind calculated. 
When the following arm of the CD crosses the sightline, it does so twice 
because of its curvature. As soon as it does so, the length of sightline 
passing through the WC7 wind increases rapidly as the orbit progresses. 
At the same time, the density of the WC7 wind traversed by the sightline 
increases as the stars approach each other. These effects combine to provide 
the rapid increase of absorption in a very short phase interval. 
The extinction in the WC7 wind to each of the intersection points was 
calculated using equation~\ref{extn} above. 

Comparison of the set of absorption vs. phase relations for different values of 
$\theta$ with the observed rise in absorption, yields $\theta = 34\pm1\degr$. 
This is smaller than the value (50\degr) derived by \citet{WVA} using different 
orbital elements, those from \citet{MM03}.
The question arises: have we really measured $\theta$ or have we measured the 
cavity in the undisturbed WC7 wind, $\theta$+$\Delta\theta$, where $\Delta\theta$ 
is the width of the shock in the WC7 wind if the structure is adiabatic? 
This is equivalent to inquiring whether the extinction in the 1.083-$\mu$m line 
occurs in the WCR as well as in the undisturbed WC7 wind. The answer is provided 
by the short-term variations in absorption in the 2008 UIST observations 
attributed to turbulence in the WCR noted above, indicating that absorption in 
the He\,{\sc i} line, and therefore the occultation, occur in the WCR and not 
(only) in the undisturbed WC7 wind.

\begin{figure}                                     
\centering
\includegraphics[width=8cm]{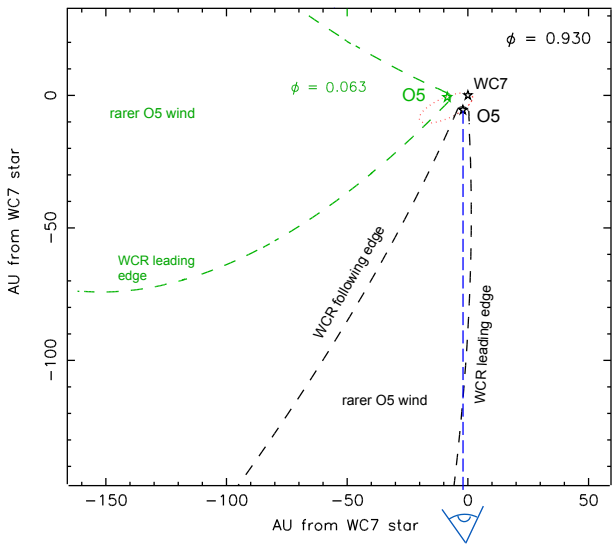}                        
\caption{Sketch of the binary and WCR configuration projected on to the plane 
of the observer and O5 star at phase 0.930 showing the of the line of sight 
to the O5 star as it was intersected by the leading arm CD (dashed line) moving 
counter-clockwise with the O5 star in this representation. This occurs much 
further from the stars than the occultation (Fig.\,\ref{Foccult}), so the 
wind density and absorption are much lower. The distance to the intersection 
increases and absorption falls with increasing phase. The curvature of the CD 
and WCR are less than at the time of the occultation because of lower transverse 
velocities. For comparison, we plot (green) the configuration at the time 
of our last observation in June 2017. The different apparent opening angles 
is a projection effect, see text.}
\label{FConfig93}
\end{figure}                                   

Besides the shocked WC7 wind on the `outside' of the CD, the WCR also includes 
the shocked O5 wind on the `inside' but, owing to the absence of 1.083-\micron\ 
absorption in the spectra of mid-O type supergiants (Section~\ref{Sintro} above) 
and the significantly higher abundance of helium in the WC7 wind, we assume 
that the observed extinction arises in the `outside' shock. 
Therefore, we can be confident that $\theta = 34\pm1\degr$ is the shock angle, 
i.e. the angle of the contact discontinuity inside the shock.
Given the size of the WCR, we can also look at the falling absorption to the 
O5 star from the beginning of our observing programme before conjunction as 
the leading arm of the CD swept past our sightline. 
At the time of our first observation, at $\phi = 0.9304$, our sightline 
to the O5 star crosses the leading edge about 120~AU from the WC7 star 
(Fig.\,\ref{FConfig93}) and this distance increases rapidly in the next few
observations. 
These distances are very much greater than that to the intersection of the 
sightline and the following edge of the WCR during the occultation, 
($\sim$ 7--10~AU, Fig.\,\ref{Foccult}), so the change in absorption with 
changing phase is very much smaller. Also shown in Fig.\,\ref{FConfig93} is 
the configuration for $\phi$ = 0.063, near the phase of our last observation. 
As the system moves from this phase to $\phi$ = 0.93, the sightline to 
the O5 star crosses the leading arm of the WCR at ever increasing distance 
from the stars, leading to ever less absorption. 
Observations of the absorption component in this phase range could 
help map the leading arm of the WCR. The 1988 and 1991 observations 
near phases 0.4 and 0.8 (Table~\ref{TCGS4}) give EWs lying between those 
at phases 0.063 and 0.93 (Tables~\ref{TGNIRS} and \ref{TUIST} respectively).

With our value of $\theta$, the sightline to the WC7 star also emerges from 
the WC7 wind briefly, near conjunction, but at a significantly greater distance 
from the WC7 star, $\sim$ 124 AU.  The lower density of the WC7 wind at this 
greater distance accounts for our not observing any significant reduction of 
the absorption component close to conjunction when the WCR, narrower than it 
is in the plane of the O5 star, crosses the sightline to the WC7 star.

We can also use our value of $\theta$ to estimate the wind-momentum ratio 
$\eta$. As noted above, the relation between $\theta$ and $\eta$ depends on 
the conditions in the shocked gas. A radiative shock gives $\eta = 0.025$ 
\citep[eqn 28]{Canto} but, as found in Section~\ref{Semission}, the variation of 
the sub-peak emission strength with the separation of the WC7 and O5 stars 
suggests that the post-shock WC7 wind was adiabatic until $\phi \simeq 0.99$, 
i.e. including the phase of the occultation from which $\theta$ was measured. 
From the relations between $\theta$ and $\eta$ of \citet{GayleyAngle}, 
the corresponding wind-momentum ratio will be smaller. 
The `characteristic angle' for an adiabatic shock applies to gas that is 
spread out beyond the contact discontinuity, so we approximate this by 
adding half the $\Delta\theta = 10\degr$ suggested by the sub-peak fitting 
Section~\ref{Semission} below to $\theta$ and derive $\eta \simeq 0.017$ for 
the wind-momentum ratio following \citet[eqn 11]{GayleyAngle}. 
The implication is that, if $\eta$ remains constant, the opening angle will 
fall when the shocks become radiative but as this geometry is an approximation, 
we assume here that the opening angle remains the same around the orbit.

\subsection{The short-term variation of absorption}
\label{Svar}

\begin{figure}                                
\centering
\includegraphics[width=8.5cm]{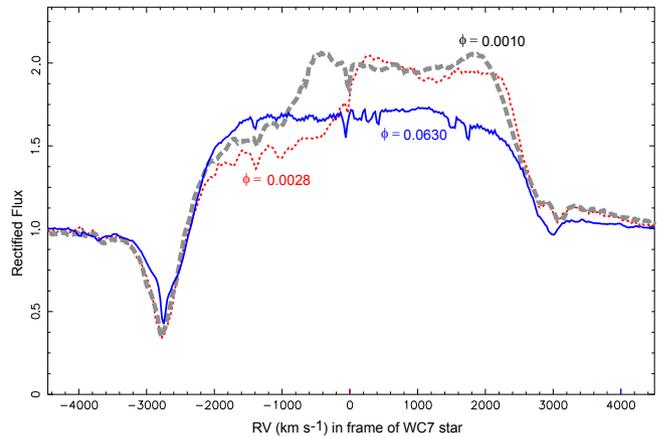}                        
\caption{Comparison of line profiles observed with GNIRS superimposed to show 
development of broad absorption between near periastron, $\phi = 0.0010$, and 
near conjunction, $\phi$ = 0.0028, (dotted line, red) where the absorption 
extended from negative velocities to $\sim$ --1000\kms\ and zero (respectively), 
eroding the emission profile. 
For comparison, we show the spectrum observed on Jun 19, $\phi$ = 0.0630, 
(blue) where the broad absorption has gone and the sub-peak emission 
at its weakest. (The differences near +3000\kms\ are not considered real but 
reflect imperfections in the correction for telluric features in the region 
of the Pa$\gamma$ line in the standard stars.)}
\label{FAbsConj}
\end{figure}                                 

Some of the closely spaced sequences of observations show development of 
short-lived maxima in extinction on a time-scale of days, e.g.: near phase 
0.992 when the extinction was rising in 2008 December (Table \ref{TUIST}); 
reaching maximum in 2016 December, between periastron and conjunction 
(WC7 star in front); 
and a narrow subsidiary maximum in the 5-day sequence of observations in 
2017 March (Table \ref{TGNIRS}) following a steady fall between phases 0.01 
and 0.03. Comparison of the profiles of the absorption feature in the higher 
resolution GNIRS observations point to the cause: absorption over a greater 
velocity range and clumpiness in the sightline attributable to instabilities such 
as those found in hydrodynamical modelling \citep*[e.g.][]{SBP}. 

The effect is greatest near periastron and conjunction when the O5 star 
and WCR are beyond the WC7 star. Profiles near these phases are compared 
with the last in our sequence of observations in Fig.\,\ref{FAbsConj}. 
They show absorption extending to near zero RV, and also broadening of the 
absorption component near --2800~\kms.
The broader absorption immediately after periastron recalls the sudden 
broadening of the absorption troughs in the ultraviolet C\,{\sc ii}, 
Si\,{\sc iv} and C\,{\sc iv} resonance line profiles at this phase 
\citep{Diah140}, to which we return below.

\begin{figure}                     
\centering
\includegraphics[width=7.5cm]{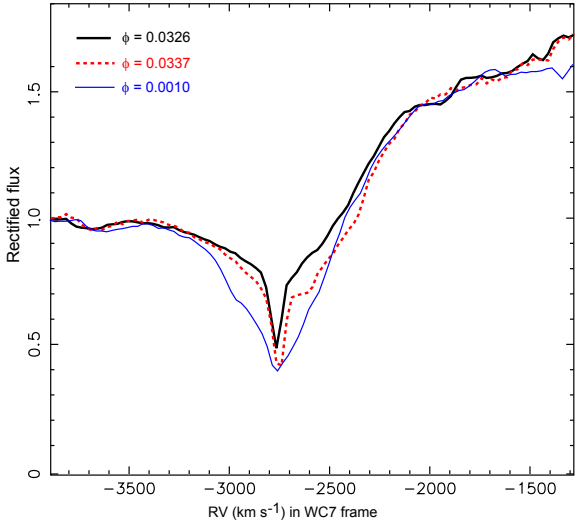}                        
\caption{Comparison of absorption line profiles observed with GNIRS -- all at 
the same instrumental resolution -- showing development of broad absorption 
attributable to turbulence in the WCR instabilities in the first two of the 
2017 March sequence of observations ($\phi \simeq 0.03$), and the even 
greater absorption observed on 2016 December 22 ($\phi = 0.0010$), very close 
to periastron.}
\label{FAbs17}
\end{figure}                                

\begin{figure}                                          
\centering
\includegraphics[width=7.0cm]{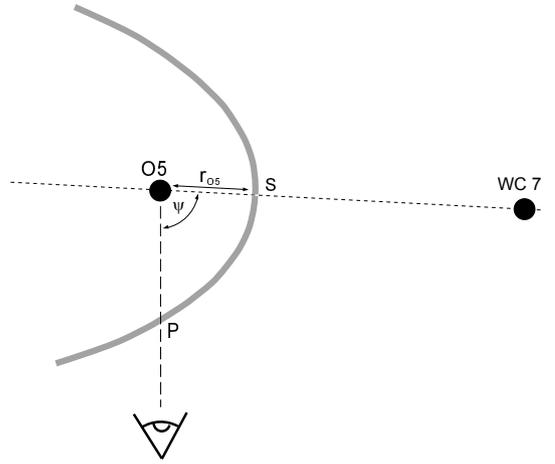}                        
\caption{Sketch of the configuration at the time of the 2017 March sequence of 
observations ($\phi \simeq 0.03$). The positions of the WC7 and O5 stars, and the 
WCR stagnation point (S) between them are marked. A thicker line shows the 
projection of the CD, which, in the absence of rotation, has cylindrical symmetry 
about the O5--WC7 axis. Our sightline to the O5 star crosses the CD at P. 
To show the relevant points, the sketch is not to scale.}
\label{Fph03}
\end{figure}

Three \ion{He}{i} profiles are compared at higher scale in Fig.\,\ref{FAbs17}. 
That observed on 2017 March 23 is the narrowest of all those observed with 
GNIRS after periastron and shows a narrow core attributable to absorption of 
the stellar continua through the undisturbed WC7 wind at its asymptotic velocity. 
The EW of the absorption (4.6\AA) is consistent with the superposition of 
the 4.6-\AA\ absorption towards the WC7 star (see above) and about 4.6\AA\ 
towards the O5 star through the WC7 wind. Three days later, 
the absorption is not only stronger but the profile has wider wings, 
extending from $\sim$ --3000 to --2400~\kms. 
The greater velocity range on its own might indicate thermal broadening, 
but the short term variation points to formation in highly turbulent dense 
clumps. The sequence of observations (Table~\ref{TGNIRS}) from the next few 
nights shows variable absorption at a comparable level, indicating the 
presence of dense structures. 
What is puzzling, however, is that the broader absorption profiles are centred 
close to the terminal velocity of the WC7 wind. If turbulence was isotropic, 
we might expect to observe velocities centred on that of the shock-compressed 
wind flowing along the CD where it is intersected by our pencil-beam sightline 
to the O5 star. The configuration at the time of the 2017 March observations 
is sketched in Fig\,\ref{Fph03}. The speed of the compressed wind along the 
CD at the intersection (P) with our sightline calculated from the O5 and WC7 
wind velocities and the thin-shell model of \cite{Canto} is $\sim$ 1600~\kms, 
giving a radial velocity $\sim$ --1130~\kms. This is far from the centre of 
the observed broadening. The difference is even greater at the time of the 
$\phi = 0.0010$ observation (Fig.\,\ref{FAbs17}) immediately after periastron, 
which shows even broader absorption. 
At this phase, the angle between the line of centres and our sightline was 
smaller, $\psi = 39\degr$, so that the intersection point P was closer to the 
stagnation point S and the speed of the compressed wind had reached only 
$\sim$ 770\kms\ at the intersection point and the component in our direction 
was only $\sim$ --260~\kms. The profile (Fig.\,\ref{FAbsConj}) does show 
extension of absorption redward, only to $\sim$ --1000~\kms, which could 
be produced in the compressed wind flowing in the WCR. At $\phi = 0.0028$, 
$\psi = 29\degr$ and the compressed wind is moving almost at right angles 
to our sightline, so its RV is $\sim$ --150~\kms. This is consistent with 
the broad absorption extending to near zero RV observed in Fig.\,\ref{FAbsConj} 
but not with the central velocity of the strongest absorption component.

Inspection of the other GNIRS spectra shows that all of them are to some 
extent affected by broad absorption outside the narrow core seen in the 
2017 March 23 spectrum (Fig.\,\ref{FAbs17}).
The broader, variable absorption is taken to be that towards the O5 star 
during the phase range when our sightline to it passes through the WCR. 
Even the June sequence near phase 0.06, about six months after periastron, 
shows a transient broad absorption feature near --1650~\kms\ on June 17,
(Fig.\,\ref{Fdynamic}), which faded over the next two nights.

In contrast to the strong profile variations seen when our sightline passes 
through the WCR, the sequence of UIST observations in 2008 June to December 
(0.92$<\phi<$ 0.98), when we view the WC7 star though its own undisturbed 
wind, show no evidence for short term variation. 
This suggests that the observed line profile variations are caused by 
high density regions or clumps in the WCR, while the undisturbed WC7 stellar 
wind is rather smooth and unclumped, at least far from periastron.

Near periastron, however, when the O5 star and WCR are beyond the WC7 star, 
the broadening of the high velocity blue-shifted P-Cygni absorption component 
suggests some large-scale disturbance from the wind from the WC7 star, perhaps 
induced by the proximity of the O5 star. It is hard to envisage a mechanism 
for this effect -- the wind of the O5 star is held close to that star by the 
WCR -- but perhaps the combination of the O5 star's continuum flux on the 
WC7 wind, coupled with the high orbital speed of the WC7 star near periastron, 
could somehow play a role.

The absorption components of the higher resolution (64 \kms) CGS4 spectra 
observed by \citet{VWA} in 2001 March are not significantly broader than 
those observed in 2000, suggesting that absorption by turbulent material 
in the WCR was not important at those times. This is in accord with the EWs, 
which show fading towards the sequence observed at a later phase in 
2017 March (Fig\,\ref{FEWphase}).

At this phase, we are sampling the WCR relatively close to the O5 star. 
To get an indication of the extent of the turbulence down-wind along the WCR, 
we can use the first of the short-lived maxima listed in Section~\ref{SAbs} 
above, the one while the absorption was rising in 2008 December. At this phase 
($\phi \simeq 0.992$), the sightline cut through the WCR between 3 and 32~AU 
from the O5 star, so that the clumps could be located anywhere in this range.

\subsection{The emission components of the 1.083-$\mu$m He\,{\sc i} line profile}
\label{Semission}

The emission profile of the 1.083-$\mu$m line in WR stars is usually very broad 
owing to its formation where the wind has attained its terminal velocity, and 
often flat-topped owing to its low optical depth. 
In colliding wind binaries, the profile can be modified by two effects: the 
emission `sub-peaks' from the shock-compressed wind flowing in the WCR, and a 
possible deficit in the underlying profile owing to missing emission from 
the cavity in the WR wind caused by the WCR \citep{SH1083} -- provided that the 
WCR lies within the region of the WC wind where the 1.083-$\mu$m emission arises. 
To investigate this, we inspect the line formation calculated with an 
appropriate PoWR atmosphere model \citep[e.g.,][]{Sander2015} for the WC7 star. 
While the detailed binary atmosphere analysis will be presented in a forthcoming 
paper, we found that most of the 1.083-\micron\ line emission is generated within 
$100\,R_\ast$. This is similar to what has been found for other WC wind models  
by \citet[WC5 star]{Hillier89} and \citet[WC8 star]{Dessart00}. More than 75 per 
cent of the line is formed within $1\,$AU and 85 per cent within a distance 
corresponding to the separation of the WC7 and the O5 at periastron. 
Hence, we do not expect a strong effect on the emission line 
caused by a cavity in the WR wind and will assume an invariant 
underlying profile for the 1.083-\micron\ emission line.
 
In order to characterise the emission sub-peaks, we need a reference spectrum 
of the undisturbed WC7 wind. All the spectra in the present programme were 
observed at phases at which the wind collisions were strong, and show at least 
some sub-peak emission. The early observations made further from periastron 
showing flat-topped profiles referred to in Section~\ref{Sintro} are unsuitable 
as sub-peak-free templates for our spectra because they have lower resolution 
or poorer signal-to-noise. Instead, a synthetic undisturbed wind spectrum was 
formed from the four (100-\kms\ resolution) UIST observations taken in 2016, 
taking their mean but replacing the fluxes in the velocity range
 --2100 to --500\kms, covering the sub-peaks, with the mean of the fluxes 
between --500 and +500\kms, omitting that at zero velocity which shows a dip 
from the photospheric absorption line in the O5 star. This is far from ideal, 
but the 1.083-\micron\ sub-peaks are, in most cases, so strong that such a 
template allows us to measure their fluxes and model their profiles without 
the introduction of significant uncertainties.

\begin{figure}                           
\centering
\includegraphics[width=8.4cm]{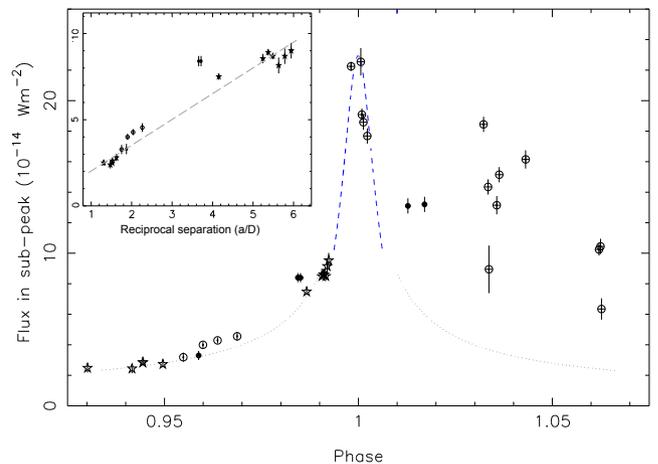}                        
\caption{Flux in the emission sub-peak from the UIST 2008 ($\star$) and 
2016 ($\odot$), CGS4 ($\bullet$) and GNIRS ($\oplus$) observations plotted 
against phase. Superimposed on the data is a dotted line representing the 
variation of the reciprocal of the separation of the WC7 and O5 stars, $D$, 
and a dashed line (blue in the on-line figure) the variation of $D^{-2}$ with 
phase, both with arbitrary normalisation. The error bars are $\pm1\sigma$. 
Inset: the $0.93<\phi<0.993$ fluxes plotted against reciprocal separation, 
$a/D$, of the stars.}
\label{Fsubphase}
\end{figure}                                   

The fluxes in the sub-peak were calculated by integrating the emission 
component, including all the features, subtracting the template spectrum, 
and converting to flux units using the continuum flux level derived above. 
They are listed in Tables~\ref{TUIST} (UIST) and \ref{TGNIRS} (GNIRS). 

The fluxes are plotted against orbital phase in Fig.\,\ref{Fsubphase}. 
Prior to periastron, the sub-peak fluxes increase steadily with phase 
whereas, after periastron, they decline very irregularly. 
Up to $\phi \simeq 0.99$, the fluxes are approximately inversely proportional 
to the stellar separation, $D$, as can be seen in the plot against reciprocal 
separation, $a/D$, in the inset, suggesting that the shocked WC wind in the 
region of the WCR where the 1.083-$\mu$m subpeak arises is adiabatic in this 
phase range, by analogy with the expected $1/D$-variation of the X-ray 
luminosity with stellar separation in such a regime \citep{SBP} and in 
accord with their expectation that the shocks in WR\,140 would be adiabatic 
for most of the orbit. 
Unfortunately, there is a gap in our temporal coverage because a spell of poor 
observing conditions prevented our getting intensive observations of the 
1.083-$\mu$m subpeaks in the critical phase range, but it is apparent that, 
closer to periastron, the flux varied more steeply with separation than as 
$D^{-1}$, as can be seen by comparison with the dashed line in the figure. 
This chimes with the demonstration by \citet{MM03} and \citet{Remi} 
that, between phases $-0.01$ and $0.01$, the flux in the sub-peak on the 
5696-\AA\ C\,{\sc iii} line varied with the separation approximately as 
$D^{-2}$. 
Taken together, these results suggest a change in conditions in the 
post-shock WC7 wind some time near $\phi \simeq 0.99$, which may be 
related to the onset of dust formation at this 
phase\footnote{Dust {\em emission} first appears at $\phi = 0.0$ but there 
is some delay \citep{WIAUC169} after the formation of sufficiently compressed 
wind for it to  flow down the WCR to be far enough from the stars for the 
grains condensing in it to survive the stellar radiation fields.} and the 
requirement of efficient cooling for this to take place \citep{Usov91}.

Immediately after periastron, until $\phi \simeq 0.003$ when there is a gap 
in our coverage, the 1.083-$\mu$m sub-peak flux appears to fade as $D^{-2}$ 
but, later, when the observations resumed after $\phi \simeq 0.03$, it was 
found to fade very irregularly, sometimes on a short time-scale, with levels 
not far below the maximum near periastron. At this phase, the binary separation 
was the same as that at $\phi = 0.97$, so that the pre-shock densities of 
the undisturbed WC7 and O5 winds and therefore the wind material available to 
be compressed in the WCR would have been the same as those at the earlier phase. 
If the formation of the sub-peak emission is by recombination, with emission 
proportional to the square of the density, formation in dense clumps, such 
as those found from the variations in the absorption component above, could
be responsible given a suitably low filling factor and the clumps remaining 
optically thin in the line. 
Throughout our sequence of observations, up to $\phi \simeq 0.063$, the flux 
does not return to the levels and dependence on stellar separation seen before 
periastron. Further observations will be needed to determine at what phase 
it does so -- and, indeed, the range in phase over which the sub-peak is 
observable. For the present, we can divide the behaviour of the emission 
sub-peak into three regimes. 
Prior to $\phi \sim 0.99$, the flux varied relatively smoothly proportionately
to $D^{-1}$ with the exception of two values near $\phi \simeq$ 0.984 observed 
in 2000 (Table~\ref{TCGS4}), which may reflect clumpiness or variation between 
cycles. Secondly, through periastron, the flux varied more steeply, possibly in 
proportion to $D^{-2}$, but the 
data are too sparse to be certain and further observations at higher cadence 
are needed to test this and check for short-term variations attributable to 
emission from clumps, such those seen in the later data. Third is the chaotic 
regime described above, where the post-shock wind appears to be very clumpy.

\subsection{Modelling the 1.083-\micron\ sub-peak.}

The observational link between colliding winds and emission-line sub-peaks 
comes from the systematic variation of their radial velocity profiles as 
the orientation of the WCR and the shock-compressed wind flowing through it 
vary around the orbit \citep{Luhrs}. 
Provided that the winds collide at their terminal velocities, the shape of 
the WCR, which is determined by the wind-momentum ratio $\eta$, does not 
change, but other geometric parameters such as the orientation and twisting 
of the WCR from orbital motion, and the velocity of the compressed wind in 
which the sub-peaks form, take up a range of values within the WCR -- as does 
the line emissivity. These phenomena have yet to be comprehensively modelled, 
but geometric models for the systematic movement of the sub-peaks in the 
spectra of CWBs during orbital motion were first developed by \citet{Luhrs} 
and since extended by \citet*{Hill02,Hill18}.  
Such models have the compressed wind moving at a constant `streaming velocity', 
$v_{\textrm{strm}}$, in a shell near the surface of a cone approximating the 
WCR. Twisting of the WCR is accommodated by giving the cone a single tilt 
angle in the orbital plane.  
The sub-peak forming region is effectively collapsed to a ring on the cone 
where conditions, including (implicitly) the emissivity are constant. 
Application of such a model, fitting the observed profiles as a function of 
phase, allows determination of quantities like $\theta$, $v_{\textrm{strm}}$, 
the orbital inclination, $i$, the tilt angle, turbulence and further 
parameters introduced to refine the model \citep{Hill18}.
Where the sub-peaks are too weak for their profiles to be determined, the 
bulk radial velocities of the compressed wind can still be modelled in a 
similar way.

\citet{Remi} applied the L\"uhrs model to the variation of the 5696-\AA\ 
sub-peak velocities within $\sim 0.01P$ of the 2009 periastron, extending 
it to allow for the rapid variation of the tilt angle around periastron by 
introducing a constant phase shift, and deriving $\theta = 39\pm3\degr$, 
$v_{\textrm{strm}} = 2170\pm100$~\kms\ and $i = 55\pm6\degr$. Similar values 
to these were derived from the previous periastron passage by \citet{MM03}.

In the case of the 1.083-\micron\ sub-peaks, we have observations of the line 
profile over a larger phase range, 0.93--0.07, which, because of the high 
eccenticity of the orbit, samples the geometry around most of the orbit, 
including both conjunctions and both quadratures. 
Because we already have values for quantities like the orbital inclination and 
$\theta$ from other observations, and can derive the flow velocity from the 
measured stellar wind velocities and $\eta$, we will not attempt to solve for 
them from the observed 1.083-\micron\ sub-peak velocities but will instead 
examine the extent to which the velocities can be recovered taking into 
account the effects which we believe may be determining them. 

Earlier studies with smaller data-sets by \citet{VWA} and \citet{WVA} modelled 
the 1.083-\micron\ sub-peak emission by considering it to arise on a cone, 
analogously to the L\"uhrs model. Here the flow velocity was taken to be the 
asymptotic velocity of the compressed wind, $v$, calculated from the WC7 and 
O5 terminal wind velocities \citep{WE2058,Diah140} following the thin shell 
wind-collision model of \citet*[eqn 29]{Canto}, which is based on the 
conservation of the momenta of the two stellar winds.  
Resolving its components $v_{\rmn{axis}}$ parallel to and $v_{\rmn{z}}$ 
perpendicular to the axis of symmetry, the emission by material flowing 
on this cone at any phase has radial velocities in the range: 

\begin{equation}
RV = v_{\rmn{axis}} \cos(\psi) \pm v_{\rmn{z}} \sin(\psi)
\label{RV}
\end{equation}

\noindent where $\psi$, the phase-dependent angle between our line of sight 
and the axis of symmetry of the WCR (equal, in the absence of orbital motion, 
to the line of centres through the WC7 and O5 stars) is defined in 
eqn~\ref{psidef} above.

\citet{VWA} showed that such a model reproduced the variations of the RV and 
velocity width of the 1.083-\micron\ sub-peak in their small data-set, for 
an opening angle $\theta = 60\degr$ and inclination $i \simeq 65\degr$. 
With addition of the 2008 UIST data, the velocity variations could be 
recovered by a similar model but with a smaller opening angle, 
$\theta = 50\degr$ \citep{WVA}, while including our new 2016--17 data, 
application of such a thin-shell model suggested $\theta$ = 53\degr.

\begin{figure}                                       
\centering                                 
\includegraphics[width=8.5cm]{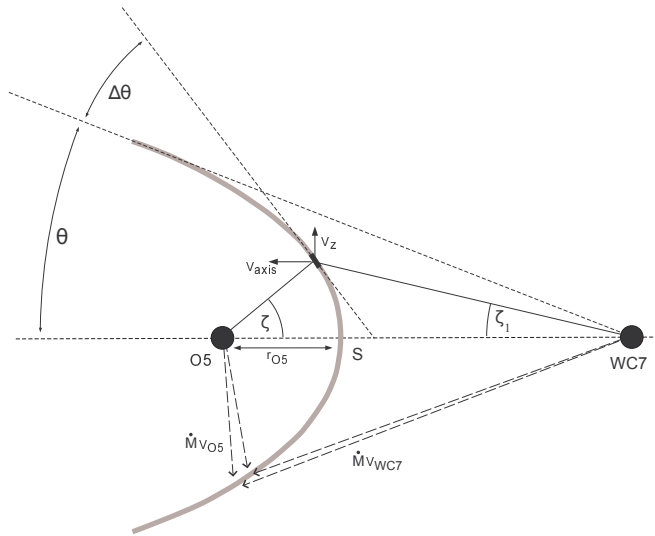}                        
\caption{Sketch of the contact discontinuity (grey) where the momenta, $\dot{M}v$, 
of the WC7 and O5 winds balance. It crosses the line of centres at the stagnation 
point (S) and has cylindrical symmetry about that line. Each point on it can be 
specified by the angles $\zeta$ and $\zeta_1$ at the O5 and WC7 stars respectively. 
In the asymptotic limit, $\zeta_1$ corresponds to the opening angle $\theta$. 
The sub-peak emission is assumed to arise in a shell outside this in a region 
of angular width $\Delta\theta$ and the flow to orginate on the CD between 
the asymptotic region and the tangent point corresponding to $\Delta\theta$.}
\label{FCantoWCR}
\end{figure}                                   

These values of $\theta$ derived from the sub-peak are significantly greater 
than that, $\theta = 34\degr$, derived above from the eclipse of the WCR. 
This difference suggests that the sub-peak emission formed some distance 
from the CD in the shocked WC7 wind in the adiabatic region of the WCR -- 
at least in the phase range when the post-shock WC7 wind was adiabatic -- 
analogous to formation in the centre of a thick mantle in the L\"uhrs model.

We now explore the ways in which the modelling can be extended to gain insights 
into  processes in the WCR when wind-collision effects are at their strongest, 
taking advantage of the range of orientations of the WCR system determined by 
the well constrained orbit. 
We are not attempting to model the likely variation of the sub-peak emissivity 
in different regions of the WCR but there are several physically motivated 
respects in which the simple geometric models can be developed; 

First, we need to consider emission from that region of the WCR where the 
compressed wind is still accelerating to its asymptotic velocity reached 
`down stream' in the region of the WCR which can be approximated by a cone. 
This follows from our observation of strongly varying absorption in the 
1.083-\micron\ profile when the sightline passes through the curved region of 
the WCR between the stars, the `shock cap' \citep{ParkinPittard}, so we must 
consider emission arising there too. We do not have a generalised relation 
for the acceleration of the compressed wind from the stagnation point or a 
relation between its velocity and the angle between its direction and the 
axis of symmetry so will use the thin-shell model of \cite{Canto}.
The condensed wind accelerates along the CD (Fig.\,\ref{FCantoWCR}, which 
follows \citet{Canto} but replaces their $\theta$ and $\theta_1$ with $\zeta$ 
and $\zeta_1$ to avoid confusion with the opening angle $\theta$), with 
velocity rising from zero near the stagnation point, $S$, to its asymptotic 
value when the angle $\zeta_1$ at the WC7 star matches the WCR opening angle 
$\theta$. 
This ties in the suggestion above that the sub-peak emission arises 
in a shell of angular thickness $\Delta\theta$ on the WC7 side of the CD 
by considering the compressed wind to arise on the CD between the tangent 
point of the $\Delta\theta$ limit (Fig.\,\ref{FCantoWCR}) and the asymptotic 
value determined by $\theta$. 
Depending on the extent along the WCR over which the emission forms, which 
can be specified in terms of the angle $\zeta$ at the O5 star, the compressed 
wind takes up a range of velocities and angles to the axis of symmetry 
instead of the single values used in previous modelling. 
We do not expect parcels of material arising from different regions of the CD 
to retain their initial velocities, which will give rise to Kelvin-Helmholz 
instabilities, but expect the average bulk velocity of the compressed wind 
to be lower than its asymptotic velocity. 

Secondly, near periastron, the CD may move close enough to the O5 star so that 
its wind has not reached terminal velocity which could cause the shock to 
weaken \citep{Sugawara140}.  
If the radius of the O5 star is comparable to those of luminous O5 
stars \citep*[$R_* \simeq 13-18~R_{\odot}$,][]{RepolustO} and its wind 
accelerated according to the $\beta$-law, 
$v(r) = v_{\infty}(1-R_*/r)^{\beta}$, with $\beta = 1$, the velocity at 
collision would be $\simeq 0.65v_{\infty}$ at periastron but closer to 
$v_{\infty}$ for most of our observations. 
The shape of the WCR is unlikely to be affected because it depends on the 
balance of the wind momenta so that, by continuity, while the stellar wind 
is still accelerating, it will have a proportionately higher density than 
if it were moving at a constant rate, thereby preserving its momentum. 
The velocity of the compressed wind, however, will be lower and vary with 
phase, which is included in the modelling.

\begin{figure}                                     
\centering                                           
\includegraphics[width=8.5cm]{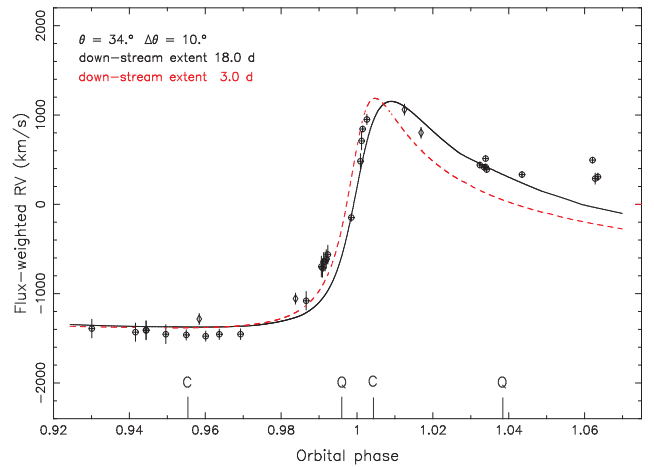}                        
\caption{Measured flux-weighted central RVs ($\oplus$ this study, $\diamondsuit$ 
RVs from spectra observed by \citet{VWA}; error bars $\pm 1~\sigma$) 
of sub-peaks compared with those calculated (line) using , the 
adopted orbital elements and phase-dependent velocity component $v_{\rmn{axis}}$ 
calculated as in the text. Short vertical lines labelled `C' or `Q' mark the 
phases of conjunctions and quadratures.}
\label{FRVc18ph}
\end{figure}                                   

Thirdly,  orbital motion will cause the axis of symmetry of the WCR to lag 
behind the line of centres through the stars. 
The relative motion of the stars causes the axis of the WCR to lag by an 
`aberration' angle determined by the ratio of the transverse velocity, 
$v_{\textrm{t}}$ calculated from the orbital motion, to the expansion velocity.  
In a long-period system like WR\,140, this angle is generally small, reaching 
only 4\fdg7 at periastron. 
Down-stream of the O5 star, the axis of symmetry and WCR are further twisted 
by the orbital motion, increasing with distance from the stars into a spiral 
structure. In this case, the degree of curvature depends on the recent 
history of the transverse velocity, $v_{\textrm{t}}$, of the O5 star and WCR 
in the orbit as well as the expansion velocity, in the same 
way as the leading and trailing arms of the WCR modelled for the occulation 
in Section~\ref{Soccult} above. Consequently, the down-stream twisting 
effect is potentially greater than the aberration and strongest some 
time after periastron. It was determined for each phase by calculating, as 
a function of down-stream distance, the difference in phase and hence that 
in the angle $\psi$ (equ. \ref{psidef}) between the axis and our sightline. 
This was added to the aberration angle for calculation of the velocities.

Before modelling the sub-peak profiles, we first examined the variation of the 
observed flux-weighted central RVs (Tables \ref{TUIST}, \ref{TGNIRS} and 
\ref{TCGS4}) with phase and compared the variation 
with that modelled using $RV = v_{\rmn{axis}} \cos(\psi)$ (cf. Eqn~\ref{RV}). 
The opening angle, $\theta$ was fixed at that determined above 
(Section~\ref{Soccult}) and the flow velocity was calculated for a series 
of incremental values of the width, $\delta\theta$ spaced by $1\degr$ 
up to the limit $\Delta\theta$ from the stellar wind velocities following 
\citet[eqn 29]{Canto}. 
To use this equation, it is necessary first to determine the position of the 
tangent point on the CD characterised by $\zeta$ and $\zeta_1$ for a given 
angle $\delta\theta$, which can be found from:

\begin{equation}
\begin{split}
\tan(\theta+\delta\theta) = 
&\big(\eta\,(\zeta -\sin\zeta \cos\zeta) + \zeta_1 - \sin \zeta_1 \cos \zeta_1\big) \\
&/ (\eta \sin^2\zeta - \sin^2\zeta_1)
\end{split}
\label{zetatheta}
\end{equation}

\noindent where $\zeta_1$, if small, is related \citep[eqn 26]{Canto} 
to $\zeta$ by:

\begin{equation}
\zeta_1 = \sqrt{15/2 (-1 + \sqrt{1 + 0.8 \eta (1 - \zeta / \tan \zeta)}}
\end{equation}

\noindent and the wind-momentum ratio, $\eta$, was calculated from our 
opening angle $\theta = 34\degr$.

This gives a series of flow velocities and angles, $\zeta$, from which we 
derive a series of $v_{\rmn{axis}} = v \cos(\theta+\delta\theta)$.
For the twisting, we determined for each phase a series of values of the angle 
$\psi$ as a function of down-stream extent. We used these and the axial 
flow velocities to derived the RVs. They are effectively volume-weighted 
but, in the absence of knowledge of the emissivity, we associated them with 
the flux-weighted RVs.

We used two fitting parameters, the angular thickness of the sub-peak 
emitting region, $\Delta\theta$, and the down-stream extent of the region 
twisted by the orbital motion. To allow for the varying size of the WCR 
around the orbit, we parameterised the latter by a constant multiple of 
the stellar separation, $D$, for simplicity; in practice, the emission 
is likely to fall off with distance from the stars as the density falls.
The data and model velocities are compared in Fig.\,\ref{FRVc18ph}, from which 
we see that the variation of the central RV with phase is recovered around 
most of the orbit -- including both conjunctions and both quadratures.  
We did not find it necessary to adopt different value of $\Delta\theta$ for 
different phase ranges and that it was fairly tightly constrained to 
$10\pm5\degr$. This width is consistent with that expected (20\degr) of the 
adiabatic WCR region corresponding to our $\theta = 34\degr$ 
\citep*{IgnaceCWB,PittardDawson} and the formation of the sub-peak within it. 
When the post-shock wind becomes radiative, the WCR is expected to become 
narrower, but we do not have enough data to test the effect of this on 
$\Delta\theta$.

The RV data were not well fit with a single value for the down-stream emission 
extent over the whole phase range; it appears that the extent is much greater 
(18~$D$) after periastron than in the $0.02P_{\textrm{orb}}$ before it, when an 
extent of 3~$D$ gives a better fit. As can be seen from Fig.\,\ref{FRVc18ph}, data 
at earlier phases do not allow us to discriminate because the transverse velocity 
and its recent history were very low. On the other hand, at phases shortly 
after periastron, when the transverse velocity had been at its maximum, the 
down-stream twisting is greatest, its effect on the sub-peak profile as 
modelled can provide a measure of the extent of the sub-peak emission.
As noted above, the sub-peak fluxes are significantly stronger after periastron, 
possibly owing to their formation in dense clumps, so that the difference in 
down-stream extent of the emission suggets that these clumps survive longer in 
the WCR than the compressed wind before periastron.

\subsection{Modelling the sub-peak profiles.}
\label{Sprofiles}  

We next sought to model the profile at the phases of our observations. 
The WCR was modelled as a series of annuli about an axis of symmetry, which 
deviates from the line of centres owing to the orbital motion as described 
above. 

The RV from any element on the annulus can be considered as the sum of three 
components: $V_1$, the projection of the flow parallel to the axis; $V_2$, 
the projection of the flow perpendicular to the axis and also to the orbital 
plane; and $V_3$, the projection of the flow perpendicular to the axis and 
within the orbital plane. They are given by:

\begin{eqnarray}
V_1 & = & v_{\rmn{axis}} \cos(\psi), \\
V_2 & = & v_{\rmn{z}} \sin(\chi) \cos(i), \ {\rmn{and}} \\
V_3 & = & v_{\rmn{z}} \cos(\chi) \sin(i) \cos(f+\omega)
\label{Vels}
\end{eqnarray}

\noindent where $\chi$ is the azimuth along the annulus on the WCR, with 
$\chi = 0$ defined as being in the plane of the orbit on the leading edge 
of the WCR, and the angle $\psi$, orbital parameters $i$, $f$, $\omega$, 
and components of the compressed wind flow $v_{\rmn{axis}}$ and 
$v_{\rmn{z}}$ are all as above. The RV components $V_2$ and $V_3$ were 
calculated for a series of angles $\chi$ around each annulus. 

The velocity components, $v_{\rmn{axis}}$ and $v_{\rmn{z}}$ depend on the 
location of the annulus on the WCR, which we specify by the angle $\zeta$ 
(Fig.\,\ref{FCantoWCR}) determined from the region of the WCR specified 
by $\Delta\theta$ as above. For the velocity of the compressed wind at each 
point on the CD specified by $\zeta$, we used Cant\'o et al.'s equation 29.

In addition, we need to consider possible lack of cylindrical symmetry in 
the emission from the annuli around the WCR axis, particularly between the 
leading and following arms of the WCR as a result of the orbital motion.
Hydrodynamical modelling of adiabatic WCRs by \citet{Lamberts12} shows 
that the outer shocks on the leading and following arms of the WCR can 
have different extents and densities. 
The effect on the observed sub-peak emission will vary round the orbit. 
Near conjunctions, in the absence of orbital motion, the compressed wind 
flowing on the leading and following arms would have similar angles to our 
sightline and hence similar projected RVs, so that any such asymmetry in 
the WCR densities would not be observable. In contrast, the effects would 
be greatest near quadrature, when the projected flows on the leading and 
following arms have opposite signs.

Weighting of the emission for azimuthal asymmetry in the WCR can be modelled 
as a function of the azimuthal angle $\chi$ defined above by 

\begin{equation}
wt(\chi) = (1 + A_1 \cos(\chi)) \times (1 - A_2\,|\sin(\chi)|\,) 
\label{asym}
\end{equation}

\noindent where the first term distributes emission between the leading and 
following arms and the second term emission in or out of the orbital plane. 
A positive value of $A_1$ favours the leading arm and a positive value of 
$A_2$ favours the orbital plane (over regions above and below it), so that, 
for example, if $A_1$ and $A_2$ have equal positive values, the product of 
the two terms loads the leading arm of the WCR, tapering out of the plane, 
and keeping the same lower weight around the rest of the annulus.

The adjustable parameters defining any model are the width, $\Delta\theta$ of 
the emitting region, which gives the lower limit on $\zeta$ defining the range 
on the CD from which the sub-peak is formed, the down-stream extension of the 
emission for the twisting of the WCR and the asymmetry parameters, $A_1$ and $A_2$. 
The relative flux at each velocity in the range $\pm4500$ \kms was calculated 
and the resulting profile was then convolved with Gaussian profiles for the 
turbulence and for the intrumental resolution, 49, 100 or 200~\kms, of the 
observed spectrum at the relevant phase to allow comparison. A range of 
different values of turbulence up to 800~\kms\ were tried when comparing the 
model and observed profiles and it was found that values below 500~\kms\ 
did not make a significant difference to the quality of the fits.  
Consequently, we adopted a uniform value of 500~\kms\ for the turbulence 
so that the influence of other parameters could more easily be seen.
The resulting profile was added to the template underlying spectrum, being 
scaled to fit the observed profile.

\begin{figure}                         
\centering
\includegraphics[width=8.3cm]{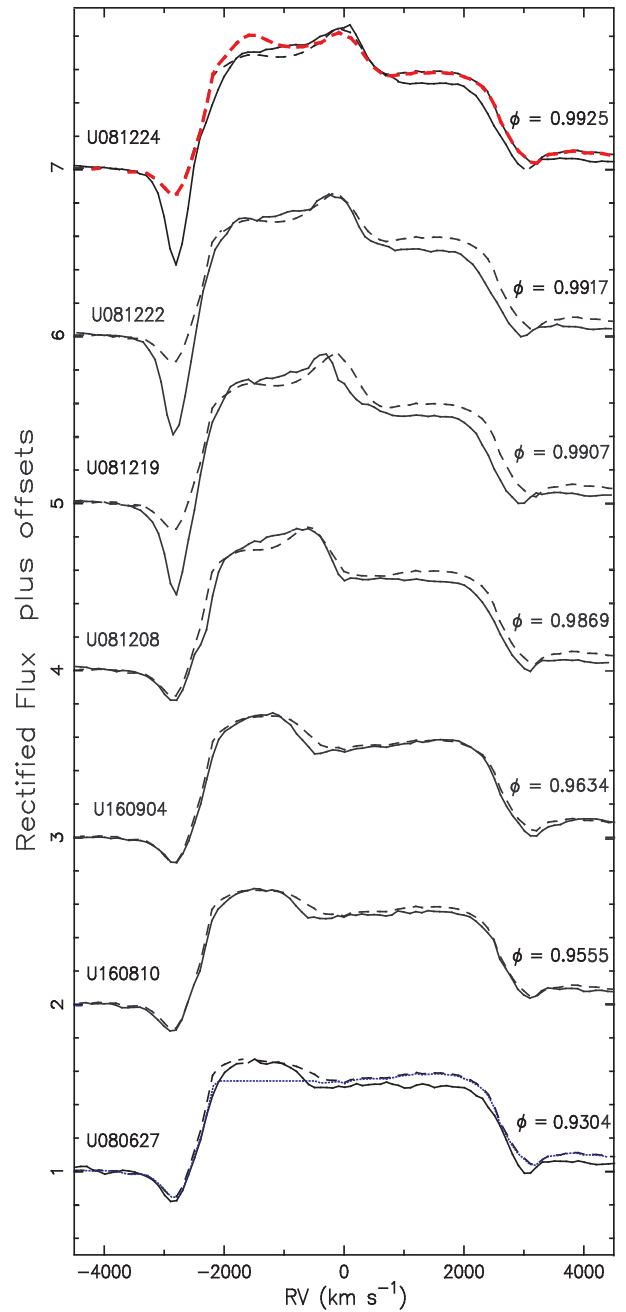}                        
\caption{Seven profiles observed with UIST from phases 0.9301 (before conjunction) 
to 0.9923 (before quadrature) labelled with dates as UYYMMDD, shifted for 
clarity, and compared with model profiles (broken lines) all calculated using 
$\Delta\theta$ = 10\degr, and a down-stream distance of $3d$, as above, but also 
allowing for azimuthal asymmetry using coefficients $A_1$ = $A_2$ = 0.5.
Overplotted (blue dotted line) on the earliest profile is the template 
spectrum without any sub-peak emission.
Another model for the 2008 December 24 (top) spectrum calculated without 
asymmetry to show the effect of the difference is shown in colour. The models 
were all convolved with Gaussian profiles for turbulence (500~\kms) and the 
instrument (200~\kms or 100~\kms for the 2008 or 2016 observations respectively).}
\label{FUmodels7}
\end{figure}                                   

We began by modelling the UIST observations (Table~\ref{TUIST}). The first 
three spectra in Fig.\,\ref{FUmodels7} bracket conjunction ($\phi$ = 0.9965, 
O5 star in front and WCR facing us) and show a strong, single, sub-peak. 
The next four spectra in Fig.\,\ref{FUmodels7} were all observed in 2008 
December and show developement of an asymmetric, rapidy broadening sub-peak. 
Our initial models of the latter using the same parameters as for the RV 
variation ($\Delta\theta$ = 10\degr, down-wind extent 3~$D$), recover the  
broadening but not the asymmetry, giving double peaks of equal height. 
The sequence ends close to quadrature ($\phi$ = 0.9965), when any asymmetry 
between the leading and following arms of the WCR would be most readily 
observable in the profiles. 
We therefore examined this effect by running models having different values 
of the asymmetry parameters, $A_1$ and $A_2$ (eqn~\ref{asym}), and found that 
$A_1$ = $A_2$ = 0.5, which has the effect of increasing the emission from 
the leading arm at the expense of that in the following arm and out of the 
plane, gave reasonable matches to the 2008 December observations. 
The effect of this inclusion is illustrated for the December 24 spectrum, 
where models with and without the asymmetry are plotted. 
Inclusion of this asymmetry for all the models in Fig.\,\ref{FUmodels7} also 
provided `infill' of the double peak of the earlier phase models through the 
re-distribution of some WCR material into the plane. 

A stronger manifestation of this asymmetry may be evident in the comparison of 
the observed and modelled {\em Chandra} HETG-MEG line profiles recently 
presented by \citet[fig 8]{Svet140}. 
The observed profiles for `Obs 1', corresponding to $\phi$ = 0.9863, close to 
that of our U081208 spectrum (Fig.\,\ref{FUmodels7}), generally show single peaks 
close to the red ends of the double-peak model profiles. 
This suggests that the region of the WCR where the \ion{Si}{xiv}, \ion{Mg}{xii} 
and \ion{Ne}{x} lines form shares the azimuthal asymmetry of that where the 
1.083-\micron\ sub-peak forms, producing stronger emission from the leading arm.
This needs to be investigated further using hydrodynamical models of the WCR and 
its emissivity \citep[cf.][]{Lamberts12}. The central velocities of the X-ray 
lines, --618 to --660~\kms\ \citep{Pollock140} are close to that (--690~\kms) 
of 1.083-\micron\ peak but the X-ray profiles fall off more sharply to the blue.

\begin{figure}                      
\centering
\includegraphics[width=8.3cm]{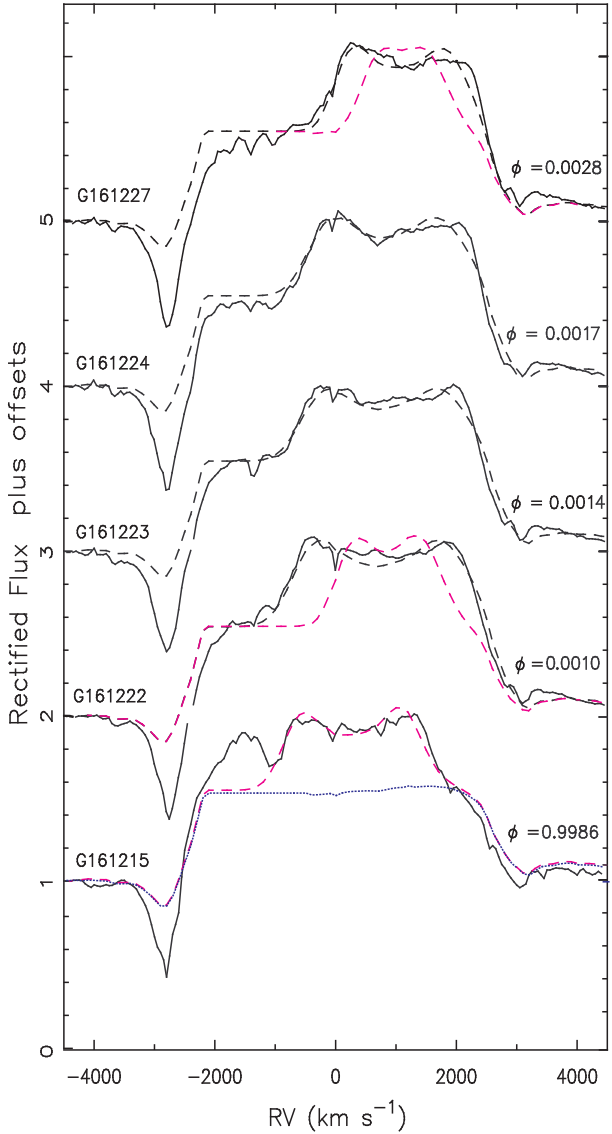}                        
\caption{Profiles observed in 2016 December with GNIRS, with dates coded GYYMMDD, 
through periastron and close to conjunction compared with models (dashed lines, 
those in red using similar parameters to those adopted for modelling the 
earlier profiles, those in black the revised parameters derived for this phase 
range). Overplotted (blue dotted line) on the earliest profile is the template 
spectrum without any sub-peak emission. The narrow absorption line near zero RV 
is taken to arise in the O5 stellar photosphere.}
\label{FGem5Dec}
\end{figure}                                   

Our next profile comes from the 2016 December 15 GNIRS observation 
(labelled G161215 in Fig.\,\ref{FGem5Dec}) at phase $\phi$ = 0.9987, an 
interval of $1.006P_{\textrm{orb}}$ after the last of the 2008 UIST spectra. 
It is quite unlike the other profiles: either there is an additional broad 
emission peak centred near RV --1700~\kms, or the emission has become very 
broad and has developed a broad absorption centred near --1550~\kms. 
As noted above, the telluric correction for this observation had to be 
taken from a spectrum observed on a different night, but there is no way 
that the broad feature could arise from a mismatch of telluric lines.
Unfortunately, there were no observations immediately before or after it, so 
we cannot trace how the features developed.  
A model profile (Fig.\,\ref{FGem5Dec}) calculated using the same parameters 
as for the earlier profiles, apart from omitting the azimuthal asymmetry, 
can match the broad emission, but not the --1700-\kms\ emission feature.
Evidently, an additional emitting stucture has come into existence within 
the WCR.

In the seven days between this last spectrum and the first of the sequence  
beginning on 2016 December 22, WR\,140 went through periastron passage. The 
last of the sequence, December 27, was observed only two days before conjunction.
At this phase, with the O5 star beyond the WC7 star and the opening of the WCR 
directed away from us, the geometry leads us to expect the central RV of the 
sub-peak to show the greatest difference from that during the $\phi = 0.9554$ 
conjunction (Fig.\,\ref{FUmodels7}) when the WCR opening was directed towards us, 
but the widths of the sub-peaks to be the same (cf. \citet{Luhrs}, \citet*{MMB96}). 
This is evidently not the case: plotted on the December 27 observed profile 
(Fig.\,\ref{FGem5Dec}) is a model (colour plot) of the sub-peak calculated using 
the same parameters as the earlier data. Although this recovered the central RV, 
it is significantly narrower than the observed sub-peak.

We therefore set out to fit the spectrum allowing the opening angle, $\theta$ and 
the flow velocities, characterised by a arbitrary multiple of the flow velocities 
for this phase calculated from the stellar wind velocities as in the previous 
modelling, to be free parameters, but retaining $\Delta\theta = 10\degr$ and the 
down-stream extent of 3$D$ for the twisting. The fitted profile with parameters 
$\theta = 50\degr$ and an arbitrary flow velocity multiple 1.3 is shown (black) in 
Fig.\,\ref{FGem5Dec}. 
Models using the same parameters give reasonable fits to the December 22--24 
sub-peak profiles (Fig.\,\ref{FGem5Dec}).  Also shown (colour) on the December 22 
profile is a model calculated using the `$\theta = 34\degr$ model' parameters 
for comparison.

The wider $\theta$ could result either from a change in the shape of the WCR 
as a whole, as determined by the wind-momentum ratio $\eta$, or from a change 
in the region of the WCR where the 1.083-\micron\ sub-peak emission forms, 
i.e. in a shell offset from the contact discontinuity by about 16$\degr$.
The observation that the sub-peak on the 5696-\AA\ \ion{C}{iii} line at the same 
phase in 2009 was {\em not} anomalously broad \citep{Remi}, suggests that there 
was no significant change in the shape of the WCR, favours the latter alternative.
We suggest that when the stars are closest, ionization of the helium by their 
radiation field restricts the formation of the \ion{He}{i} emission in the 
inner regions of the WCR. 
It is not clear when this change came about; it is possible that the --1700\kms\ 
flow observed in the December 15 spectrum was the beginning of the displaced 
plasma flow in the `following' arm of the WCR judging from the sign of the RV. 
We also tested the possibility that the additional ionization could force the 
\ion{He}{i} emission down-stream in the WCR, which would show up in the profiles 
through requiring a greater down-stream twisting length, but modelling did not 
support this possibility. We suggest that the enhanced flow velocity is also a 
consequence of the higher stellar radiation field in this phase range.

\begin{figure}                           
\centering
\includegraphics[width=8.3cm]{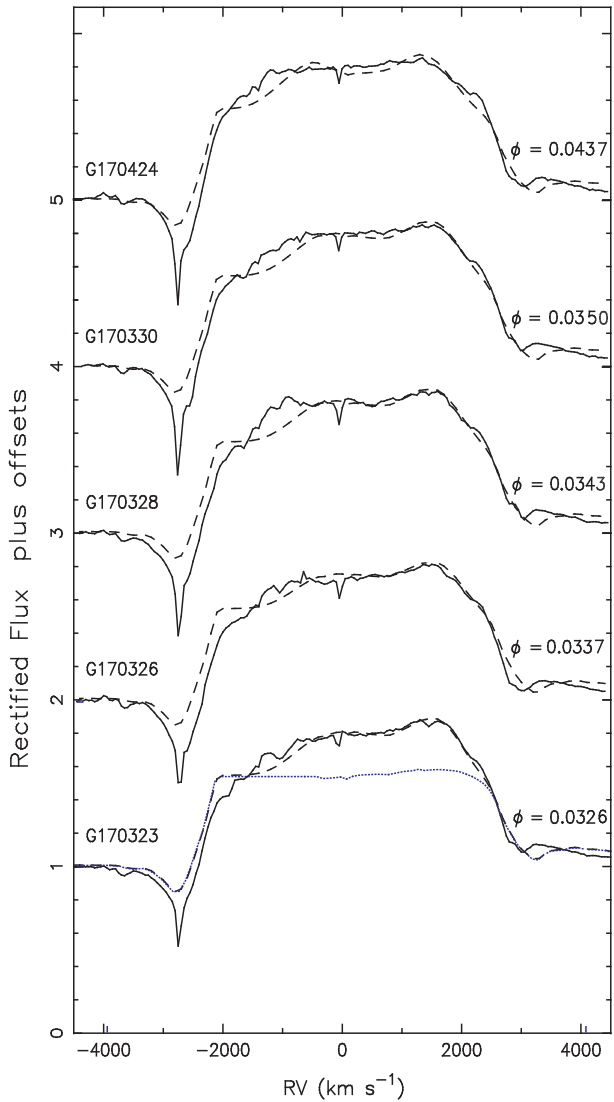}                        
\caption{Profiles observed in 2017 March--April with GNIRS, with dates coded 
GYYMMDD, compared with models (dashed lines). Overplotted (blue dotted line) 
on the earliest profile is the template spectrum without any sub-peak emission.
The narrow absorption line near zero RV is taken to arise is the O5 stellar 
photosphere.}
\label{FGem17Mar}
\end{figure}                                   

After the December sequence of observations, there was an interval of almost 
three months before the next sequence of spectra in 2017 March, followed
by one in April, shown in Fig.\,\ref{FGem17Mar}. The last two spectra bracket 
orbital quadrature, when the WCR would have been viewed side-on and the 
RV amplitude greatest -- accounting for the breadths of the sub-peaks. 
The sub-peak models shown use the same parameters as for the pre-periastron 
data with only the down-stream distance for orbital twisting increased to 
18~$D$, as derived above from the variation of central velocity with phase. 
Evidently, the anomalous broadening of the December 22--27 spectra ascibed to 
an offset of the sub-peak emission from the contact discontinuity has ceased.
The March--April sequence shows the gradual flattening of the sub-peak profile 
as the emission fades. 
This continued in the final spectra observed in June, which show a weak, flat 
emission and are not modelled because they are not very different from the 
template profile devised above.

\section{Comparison with observations at other wavelengths}
\label{SComp}

The absorption component of the 1.083-$\mu$m profile is the superimposition of 
those formed in the sightlines to the WC7 and O5 stars. Because the sightline 
to the WC7 star always passes through at least part of its own wind closest to 
the star, where the density is highest, we assume that component of the observed 
absorption to be constant and assign all of the variation observed to varying 
absorption along the sightline to the O5 star. After $\phi$ = 0.986, when the 
sightline to the O5 star starts passing through the shocked WC7 wind in the 
WCR, we observe strong and variable absorption, which reaches a maximum shortly 
after periastron passage. 

The X-ray absorption (Pollock et al., in preparation) also reaches a maximum 
at conjunction, just after periastron, when the WCR and X-ray source are 
beyond the WC7 star and suffer the greatest absorption. 
We have already drawn attention (Section~\ref{Soccult} above) to a similarity 
in the forms of the increase towards maximum of the X-ray hardness ratio in 
the {\em RXTE} PCA data, a proxy for absorption, and the 1.083-\micron\ profile 
absorption: both showing a pause near $\phi \simeq 0.995$. 

Variations in the WCR also appear to be responsible for rapid changes seen in 
the ultraviolet spectrum. 
The sequence of {\em IUE} spectra covering almost a whole orbit, including the 
1993 periastron passage, observed by \citet{Diah140} show sudden strengthening 
and broadening of the \ion{C}{ii}, \ion{Si}{iv} and \ion{C}{iv} resonance lines 
between phases 0.96 and 0.012 (on their elements; the phases on those used here 
are very similar). 
The deep absorption trough of the Si\,{\sc iv} 1394,1493-\AA\ 
doublet became saturated and broadened from 400~\kms, to 1200~\kms, then 
extending from --3200 to --2000~\kms, while that of the C\,{\sc iv} 
1548,51-\AA\ doublet behaved similarly. 
Both profiles took a long time for the troughs to recover their `quiescent', 
pre-periastron breadths -- until phases 0.3 and 0.6 respectively. 
If these effects arise in the sightline to the O5 star through the WCR, which 
seems probable,\footnote{Both profiles also show broad absorption extending 
to +3400 and +2000~\kms, which may have formed in the pre-collision WC7 wind 
as the system was close to conjunction with the WCR beyond the WC7 star at 
this phase.} they demonstrate that the WCR takes a long time to recover from 
the periastron passage.

The \ion{He}{i} 1.083-\micron\ sub-peak emission shows a very similar effect, 
varying smoothly before periastron and very irregularly after it. 
The sub-peak fluxes were significantly greater than those before periastron 
at the same stellar separations (e.g. at $\phi$ 0.965 and 0.035), when the 
pre-shock wind densities and hence material available to the WCR, would have 
been the same. 

\begin{figure}                                
\centering
\includegraphics[width=8.3cm]{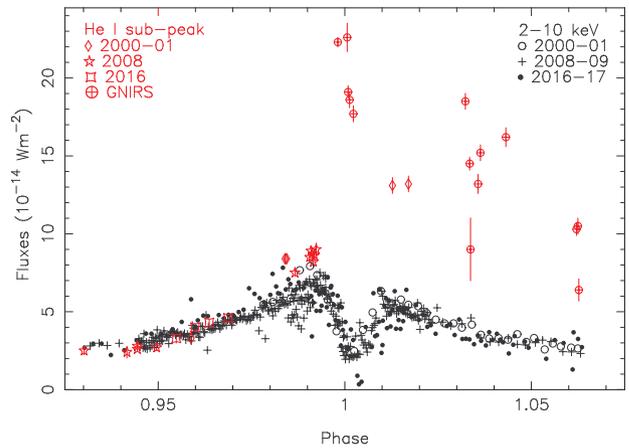}                        
\caption{Comparison of the absorption-corrected 2--10-keV X-ray and 
1.083-\micron\ sub-peak fluxes (colour) plotted against phase.}
\label{FHevsXray}
\end{figure}                                 

The asymmetry about periastron of the intrinsic non-thermal radio emission may 
be related. \citet{WhiteBecker} observed WR\,140 around its orbit at 2, 6 and 
20~cm and were thus able to derive the variation of both the intrinsic 
non-thermal emission and the circumstellar free-free absorption with phase. 
The intrinsic 2-cm non-thermal emission varied from $\sim$ 2mJy near periastron,  
reaching a maximum at $\phi \sim 0.7$; the flux density at $\phi \simeq 0.8$ 
was three times that at $\phi \simeq 0.2$. The authors suggested a model in 
which the wind of WR\,140 was flattened into a disk, but this was not supported 
by spectropolarimetry \citep*{HHHspol}, which showed no line effect -- so the 
mechanism for the variation of the intrinsic non-thermal emission remains an 
open question.

Another manifestation of asymmetric behaviour about periastron can be seen in 
the optical photometry: the $UBV$ magnitudes in 2001 after periastron ($\phi$ 
= 0.020--0.055) show dips attributed to formation of clumps of dust in the 
line of sight \citep{MM03} whereas the photometry before periastron did not 
show this phenomenon. This could be related to the clumps in the WCR after 
periastron suggested as the cause of the strength and variation of the 
1.083-\micron\ sub-peak emission in the same phase range rather than to 
the substantial dust clouds formed each periastron passage. Although their 
strong IR `excess' emission is observed only after periastron, this does not 
indicate any asymmetry about periastron of the wind-collision process, but 
reflects the prolonged cooling of the newly formed dust as it moves 
away from the stars \citep{W90,Dust140}.

The flux (Tables \ref{TUIST} and \ref{TGNIRS}) in the sub-peak on the \ion{He}{i}
profile is a significant source of cooling for the shock-heated material,
reaching a maximum of $2.3(\pm0.1)\times10^{-13}$~Wm$^{-2}$ at periastron. 
At maximum, the sub-peak on the 5696-\AA\ \ion{C}{iii} line had an EW of 
10.9~\kms \citep{Remi}, which can be converted to an integrated flux of 
$1.9 \times 10^{-14}$~Wm$^{-2}$, corrected for interstellar reddening. Other 
lines in the visible also show sub-peak emission (e.g. 5876-\AA\ \ion{He}{i}).
In the phase range $\phi$ = 0.93--0.97, the sub-peak flux was very similar 
to the absorption-corrected 2--10-keV X-ray flux (Fig.\,\ref{FHevsXray}), 
increasing as $D^{-1}$ and was an approximately equal contributor to the 
cooling of the shock. The sub-peak flux continued increasing at this rate until 
$\phi \sim 0.99$ (Section~\ref{Semission} above), after which it increased 
more quickly while the X-ray flux fell below the $D^{-1}$ dependence and 
went through a minimum close to conjunction, as discussed by Pollock et al.
After periastron, the X-ray flux recovered its earlier $D^{-1}$ dependence 
near $\phi$ = 0.02 whereas the 1.083-\micron\ sub-peak emission remained 
strong and very variable.

\section{Conclusions}
\label{SConclude}

New observations of the He\,{\sc i} 1.083-$\mu$m line around the times of the 
2008 and 2016 periastron passages showed a strong and variable P~Cygni profile.
Both emission and absorption components of the profile are powerful diagnostics, 
of very different scope. The emission comes from the system as a whole, both 
stars and WCR, whereas the absorption samples a tiny part of the system along 
two pencil beams, whose varying positions are well known from the orbit.
The strength of the absorption component showed a sharp increase at 
$\phi$ = 0.986 as the `following' arm of the WCR crossed our sightline to 
the O5 star, allowing us to set a tight limit on the opening half opening angle 
of the contact discontinuity in the WCR: $\theta = 34\degr\pm1\degr$. 

Particularly near and after periastron, when our sightline to the O5 component 
crossed the WCR, the strength and breadth of the absorption component varied on 
a short time-scale, suggesting turbulence and instabilities in the WCR, as 
expected theoretically e.g., \citet{SBP,WalderFolini}. The central velocity of 
the absorption, however, was not consistent with the expected velocity field 
in the WCR while relation to the wind of the WC7 star was also problematic. 
This remains a conundrum to be resolved.

An emission sub-peak was visible on top of the normally flat-topped emission 
profile from our earliest observation at phase 0.93. Until phase $\sim 0.99$, 
its flux was approximately proportional to the inverse of the stellar 
separation, $D$, as expected from an adiabatic post-shock wind. 
Between phases 0.99 and 0.01, the variation with separation was steeper, 
nearer to being proportional to $D^{-2}$, suggesting increased cooling of the 
the shocked wind, consistent with the condensation of dust (which requires 
efficient cooling) in this short interval. 
Thereafter, the fading of the sub-peak emission was very irregular but 
it was always significantly stronger than that at the corresponding 
stellar separations before periastron. As the amounts of stellar wind material 
available for compression in the WCR, which depend on stellar separation 
through the undisturbed wind densities, would have been the same, we suggest 
that the extra, variable emission was caused by the formation of clumps in 
the shock-compressed wind. 
Our early observations found the sub-peak flux to be approximately equal to 
the X-ray flux but, after $\phi \simeq 0.97$, the sub-peak flux exceeded 
the X-ray flux and became the major source of cooling of the shock.

New geometric models for the profiles of the sub-peaks have been developed. 
They allow for emission from the region of the WCR extending $\Delta\theta$ 
on the WC7 wind side of the contact discontinuity, which corresponds to a 
region on the CD where the shock-compressed wind is still accelerating from 
the shock apex to its asymptotic value down-stream where the WCR can be 
approximated by a cone. Consequently, the flow has a range of velocities 
and angles to the axis of symmetry. The models also allow for the twisting 
of the flow down-stream caused by orbital motion and for the occurrence 
of the wind collision so close to the O5 star that its wind could not have 
achieved its terminal velocity if it accelerated according to a $\beta$-law. 
Both these effects vary around the orbit. 
It was possible to recover the variation of the flux-weighted RV of the 
sub-peak emission over the phase range 0.93 -- 1.06 which, because of the 
high eccentricity of the orbit, includes both the conjunctions and both 
the quadratures thereby sampling practically the whole orbital geometry, with 
a a model based on the occultation-determined opening angle $\theta$ = 34\degr, 
and flow velocities calculated from the stellar wind velocities following 
\cite{Canto}. Adjustable parameters were the flow thickness, $\Delta\theta$, 
and the down-stream extent over which emission from the twisted compressed 
wind needed to be taken into account, expressed as a multiple of the stellar 
separation, $D$. We found all the data could be fit with $\Delta\theta = 10\degr$. 
Prior to phase $\sim 0.01$, twisting of the WCR for 3$D$ down-stream was 
indicated; subsequently, in the latter 
phase range in which the fluxes suggested the post-shocked wind was heavily 
clumped, modelling the effect of WCR twisting required an extent of 18$D$, 
suggesting survival of the clumps a significant distance down-stream.

Fitting the observed profiles revealed different regimes in three different 
phase ranges. Up to $\phi = 0.9925$, the profiles could be fit using the 
same parameters as for the phase-dependence of the RV with one refinement: 
allowance was made for azimuthal asymmetry of the emission about the WCR 
axis, favouring the `leading' arm of the WCR. This may also explain the 
difference between published observed and modelled profiles of X-ray lines 
observed with {\em Chandra} in this phase range. Certainly, the comparison 
of profiles of the sub-peak and X-ray lines observed contemporaneously can 
be expected to yield fresh insights to the WCR phenomenon.
Closer to periastron, modelling showed that the sub-peak emission came from 
a region characterised by a larger opening angle, suggesting formation in 
a shell offset from the contact discontinuity, possibly because of ionization 
of the helium by the intense stellar radiation field. Subsequently, from 
$\phi \simeq 0.03$, the profiles could again be fitted by the parameters used 
for the pre-periastron data, suggesting that the WCR and location of the 
sub-peak emission had recovered from the disruption of periastron passage. 
These profiles did not suggest greater sub-peak emission in the leading arm 
of the WCR, perhaps because it was moving into less dense regions of the 
stellar winds. 

These geometric models take no account of the variation in sub-peak emissivity 
in the WCR and its variation around the orbit, all which need to be modelled 
to exploit the power of the 1.083-\micron\ profile as a diagnotic of the WCR.

Further observations of the profile are also called for, earlier in phase than 
our first observation to track the reduction in absorption to map the leading 
arm of the WCR, to time appearance of the sub-peak and to get a better 
template spectrum for defining the sub-peak emission when it is weak, and, 
particularly, around periastron to track the rapid changes shown by our patchy 
coverage, such as the dependence of the flux on stellar separation, $D$, the 
development of the --1700-\kms\ feature at $\phi$ = 0.9987 (if it recurs 
periodically) and the subsequent broadening of the sub-peak in the approach to 
conjunction.
The 1.083-\micron\ \ion{He}{i} profile has proved to be a powerful diagnostic 
of the colliding winds in WR\,140 and has the potential to reveal much more.

\section*{Acknowledgments}
We would like to thank UKIRT and Gemini Service Observing astronomers for 
obtaining the spectra for this study.
Prior to November 2014, UKIRT was operated by the Joint Astronomy Centre, Hilo, 
Hawaii, on behalf of the U.K. Science and Technology Facilities Council. 
When the 2016 observations were acquired, UKIRT was supported by NASA and 
operated under an agreement among the University of Hawaii, the University 
of Arizona, and Lockheed Martin Advanced Technology Center; operations were 
enabled through the cooperation of the East Asian Observatory.
Based also on observations obtained at the international Gemini Observatory, a 
program of NSF's NOIRLab, which is managed by the Association of Universities 
for Research in Astronomy (AURA) under a cooperative agreement with the 
National Science Foundation. on behalf of the Gemini Observatory partnership: 
the National Science Foundation (United States), National Research Council 
(Canada), Agencia Nacional de Investigaci\'{o}n y Desarrollo (Chile), 
Ministerio de Ciencia, Tecnolog\'{i}a e Innovaci\'{o}n (Argentina), 
Minist\'{e}rio da Ci\^{e}ncia, Tecnologia, Inova\c{c}\~{o}es e 
Comunica\c{c}\~{o}es (Brazil), and Korea Astronomy and Space Science 
Institute (Republic of Korea).
It is a pleasure to thank Ken Gayley for a helpful referee's report.
AFJM is grateful to NSERC (Canada) for financial aid. 
PMW is grateful to the Institure for Astronomy for continued hospitality 
and access to the facilities of the Royal Observatory Edinburgh.

\section*{Availability of data}
The data underlying this article are available at the Canadian Astronomy 
Data Centre ({\tt https://www.cadc-ccda.hia-iha.nrc-cnrc.gc.ca/en/}) or 
will be shared on reasonable request to the corresponding author.

\bibliographystyle{mnras}
\bibliography{He2019}           

\bsp	
\label{lastpage}

\end{document}